\documentclass[pre, floatfix, nofootinbib, letterpaper, preprint]{revtex4}
\usepackage{float}
\usepackage[caption=false]{subfig}
\usepackage{amssymb, amsmath}
\usepackage[colorlinks,citecolor=blue,urlcolor=blue]{hyperref}
\usepackage{color}

\usepackage{graphicx}
\newcommand{\InsertFig}[4]
{\begin{figure}[h!t]
         \includegraphics[width=#4\columnwidth]{./#1}
       \caption{{\footnotesize  #2}
       \label{fig:#3}}
\end{figure}}

\newcommand{\InsertFigTwo}[5] {
\begin{figure*}[htb]
       \centerline{
         \includegraphics[width=#5\columnwidth]{./#1}
         \hskip 0.5in
         \includegraphics[width=#5\columnwidth]{./#2}
       }
       \caption{{\footnotesize  #3}
       \label{fig:#4}}
\end{figure*}}

\newcommand{\tc}[1] {\multicolumn{1}{c|}{#1}}

\newcommand{\Tr}{\mathop{\rm Tr}}
\newcommand{\bR}{{\mathbb{ R}}}

\newcommand{\bZ}{{\mathbb{ Z}}}
\newcommand{\bN}{{\mathbb{ N}}}

\newcommand{\Fix}[1]{\mathop{\rm Fix}({#1})}

\newcommand{\Eq}[1]{(\ref{eq:#1})}
\newcommand{\Sec}[1]{\S \ref{sec:#1}}
\newcommand{\Fig}[1]{Fig.~\ref{fig:#1}}
\newcommand{\Tbl}[1]{Table~\ref{tbl:#1}}

\begin{document}

\title{Statistics of the Island-Around-Island Hierarchy in Hamiltonian Phase Space}
\author{Or Alus}
\email{oralus@tx.technion.ac.il}
\author{Shmuel Fishman}
\email{fishman@physics.technion.ac.il}
\affiliation{Physics Department\\ Technion--Israel Institute of Technology\\ Haifa 3200, Israel}
\author{James D. Meiss}
\email{james.meiss@colorado.edu}
\affiliation{Department of Applied Mathematics \\ University of Colorado, Boulder, Colorado 80309-0526 USA}

\date{\today}

\begin{abstract}
The phase space of a typical Hamiltonian system contains both chaotic and regular orbits, mixed in a complex, fractal pattern. One oft-studied phenomenon is the algebraic decay of correlations and recurrence time distributions. For area-preserving maps, this has been attributed to the stickiness of boundary circles, which separate chaotic and regular components.  Though such dynamics has been extensively studied, a full understanding depends on many fine details that typically are beyond experimental and numerical resolution. This calls for a statistical approach, the subject of the present work. We calculate the statistics of the boundary circle winding numbers, contrasting the distribution of the elements of their continued fractions to that for uniformly selected irrationals.  Since phase space transport is of great interest for dynamics, we compute the distributions of fluxes through island chains. Analytical fits show that the ``level'' and ``class'' distributions are distinct, and evidence for their universality is given.
\end{abstract}

% \noindent
\pacs{XXXXXX}
\vspace*{1ex}
\noindent
% Keywords:

\maketitle

%%%%%%%%%%%%%%%%%
%% Introduction
%%%%%%%%%%%%%%%%%
\section{Introduction}\label{sec:intro}

Area-preserving maps constitute the simplest models of chaotic behavior in conservative dynamics. Such two-dimensional maps typically arise through a Poincar\'e section of a two degree-of-freedom Hamiltonian system. In virtually all physical applications where chaos arises in classical mechanics, the phase space consists of a mixture of both regular and chaotic orbits. A typical phase space, for the H\'enon map studied in this paper, is presented in \Fig{PhaseSpace7x7}. While there are many rigorously known properties of such maps (e.g. Poincar\'e-Birkhoff theory, KAM theory, Aubry-Mather theory, etc.), the striking numerical observation of an infinite hierarchy of regular islands embedded in a chaotic sea of nonzero measure \cite{Berry1978, Tabor1989}---has resisted rigorous approaches. Orbits in this chaotic ``fat fractal'' \cite{Umberger1985} have a highly irregular motion and stick for long times in the neighborhood of the boundaries of islands. This leads to an observed algebraic decay of correlation functions, Poincar\'e recurrences and survival times---that is, to distributions of the form
\begin{equation}\label{eq:AlgebraicDecay}
	P(t) \sim t^{-\gamma} .
\end{equation}
The recurrence time exponent is connected with the exponent of the mean square displacement $\left\langle (\Delta x(t))^{2}\right\rangle \sim t^{\beta}$, by
$\gamma+\beta=3$ \cite{Venegeroles2009}.

There has been much controversy over \Eq{AlgebraicDecay}
 \cite{Karney1983, Chirikov1984, Hanson1985, Meiss1986a, Zaslavsky1997, Zaslavsky2000, Weiss2003, Zaslavsky2005, Cristadoro2008, Venegeroles2009, Ceder2013}: is the decay really asymptotically a power law? If it is, does the value of $\gamma$ depend upon the specific map, its parameters, or the location in phase space?

One explanation for algebraic decay is a simple model based on a hierarchical
Markov chain or tree \cite{Hanson1985, Meiss1986a} using the transport theory of
MacKay, Meiss and Percival \cite{MacKay1984}. The point is that each island embedded
in a chaotic sea corresponds to a nested set of invariant circles surrounding a point on an
elliptic periodic orbit. The outermost of these circles, the ``boundary circle,''
is sticky because it is enclosed by remnant invariant circles---``cantori.''(see \Fig{PhaseSpace7x7}) The flux of chaotic trajectories through these cantori limits to zero approaching the boundary \cite{MacKay1984}. Moreover, each island is generically surrounded by other islands that correspond to additional (librational) elliptic periodic orbits formed where the winding number around the first orbit is rational (see e.g., \Fig{PhaseSpace7x7})---this is the Poincar\'e-Birkhoff scenario. If some of these islands are outside the boundary circle, they too are in the chaotic sea. The boundary circles of these new islands are also enclosed by sequences of cantori with fluxes approaching zero.

While there are potentially many islands embedded in a chaotic sea (for example, islands formed by a saddle-center bifurcation \cite{Tabor1989}), the simplest model keeps only the islands in the Poincar\'e-Birkhoff hierarchy; this gives the binary, Markov tree model of Meiss and Ott \cite{Meiss1986a}, sketched in \Fig{SelfSimilarTree}. The sequence of islands that correspond to successive rational approximants of the boundary circle winding number correspond to one branch of this tree and are labeled by ``level'' see \Fig{building_the_tree}. Between two neighboring levels there is a cantorus of minimal flux, and successive minimal flux cantori bound a state on the Markov tree.  A second branch of the tree is formed by the hierarchy of cantori that surround the main island at a given level. This corresponds to incrementing the ``class.'' This construction gives a binary tree (see e.g., \Fig{building_the_tree}): transitions to the right correspond to incrementing the level and thus approaching a fixed boundary circle, and transitions to the left correspond to incrementing the class, i.e., orbiting smaller island in the hierarchy.

The simplest situation corresponds to a self-similar tree.
The transition rates are then determined using the MacKay-Greene renormalization
theory \cite{MacKay1983} for the levels, and by an analogue of the Feigenbaum period-doubling
scenario for classes \cite{Meiss1986}. In this case, $\gamma$ can be
obtained analytically in the Markov approximation \cite{Hanson1985,Meiss1986a}.
A major disadvantage of this method is that the assumed self-similar structure of
islands-around-islands occurs only at specially chosen parameter values for a given
map and is only asymptotic;
moreover, the level and class self-similarities do not occur simultaneously in one-parameter families of maps.

Nevertheless, observed algebraic decay rates appear to exhibit some degree of universality \cite{Venegeroles2009}. This is, however, a delicate numerical problem since chaos, by its very definition, precludes accurate, long computations, and the observations show that the decay rates fluctuate with increasing time even up to the longest times, say $10^{10}$ iterations, that have been computed. In recent years a statistical approach has been introduced in order obtain a universal decay rate \cite{Cristadoro2008, Ceder2013}. This approach is based on the idea
that the dynamics in different regions of phase space will be described by different scalings, and in some sense this is equivalent having an ensemble of Markov models in which the transition rates are chosen randomly from some distribution. The idea is that the long-time statistics is governed by orbits that stick near small islands in various regions of phase space. The detailed structure of these islands depends upon location. Thus one can consider the long time statistics to be governed by an ensemble of localized structures.

For example, Cristadoro and Ketzmerick \cite{Cristadoro2008} showed that $\gamma$ does not depend on the realization but rather on the probability distribution of transition rates. In their work this distribution was chosen
in an ad hoc way, and to be the same for level and class transitions. The paper
\cite{Cristadoro2008} used a specific distribution, namely
a uniform one. If there is some universal distribution, then one expects that
this explains the universality of $\gamma$ for different maps and parameter values.
In the present work we provide evidence that such a
distribution does exist for the H\'enon map.

Ceder and Agam \cite{Ceder2013} assumed that the transition rates are effectively
self-similar, but added white noise fluctuations. They showed that this causes
fluctuations in $\gamma$ that do not decay even
for very long times.

We start in \Sec{Preliminaries} by reviewing some of the properties of area-preserving maps,
and in particular of the model that we will use, H\'enon's quadratic area-preserving map.
In \Sec{Numerical}, we discuss numerical methods to compute periodic orbits of various rotation numbers.
In \Sec{BoundaryCircles} we will extend the work of Greene, MacKay and Stark
\cite{Greene1986} on the distribution of continued fraction elements for boundary circles, showing that
this distribution appears to be universal over variations
in the parameter of the H\'enon map, as well as over classes
of islands.
Finally, in \Sec{Flux} we  compute a distribution of scaling factors for the
flux through periodic orbits in the islands-around-islands hierarchy. We show
that the distribution has two forms, depending upon whether the ratio is computed
for fluxes through successive levels that approach a given boundary circle, or
for fluxes between successive classes of islands.
%%%%%%%

%%%%%%%%%%%%%%%%%
%% Preliminaries
%%%%%%%%%%%%%%%%%
\section{Preliminaries}\label{sec:Preliminaries}

%%%%%%%%%%%%%%%%%
%% Area-Preserving Maps
%%%%%%%%%%%%%%%%%
\subsection{Area-Preserving Maps}\label{sec:APMaps}

A map $T$ is a discrete-time dynamical system, defining a trajectory
by the iteration $\vec{x}\,' = T(\vec{x})$ for a point $\vec{x} \in \bR^n$.
An orbit of $T$ is a sequence $\{\vec{x}_t : \vec{x}_{t+1} = T(\vec{x}_t), t \in \bZ\}$.
The map  preserves volume if
\begin{equation}\label{eq:area preserving}
	\det(DT)=1,
\end{equation}
where $DT$ is the Jacobian of $T$. For the two-dimensional case,
letting $\vec{x} = (x,y)$, the Jacobian is the $2\times 2$ matrix
\[
%	(\begin{array}{c} y'\\ x' \end{array})
%       =T(\begin{array}{c}y\\x \end{array}) \\
	DT =\left(\begin{array}{cc}
		\frac{\partial x'}{\partial x} & \frac{\partial x'}{\partial y}\\
		\frac{\partial y'}{\partial x} & \frac{\partial y'}{\partial y}
\end{array} \right) .
\]
Since this matrix has determinant one, its eigenvalues form a reciprocal pair $\lambda_1\lambda_2 = 1$. An orbit $\{\vec{x}_0,\vec{x}_1, \ldots, \vec{x}_{q-1}\ldots \}$ has period-$q$, when $\vec{x}_q = \vec{x}_0$, or equivalently if $T^q(\vec{x}_0) = \vec{x}_0$.
The eigenvalues of the Jacobian $DT^{q}(\vec{x_0})$ determine the linear stability of the orbit.
The Jacobian of the composition can be computed using the chain rule
\[
	DT^{q}(\vec{x}_0)=\prod_{t=0}^{q-1}DT(\vec{x_{t}}) .
\]
For the area-preserving case, John Greene encoded stability in a convenient index he called
the ``residue'' \cite{Greene1979},
\begin{equation}
	R=\tfrac{1}{4}(2-\Tr(DT^{q})) .
\end{equation}
When $R<0$ the fixed point is hyperbolic ( $0 < \lambda_1 < 1 <\lambda_2$), when $0<R<1$ it is elliptic ( $|\lambda_i| = 1$)
and when $R>1$ it is hyperbolic with reflection ( $\lambda_i < 0$). When the fixed point is elliptic, it has eigenvalues $e^{\pm 2\pi i \omega_0}$, and the local rotation frequency $\omega_0$ is related to the residue by
\begin{equation}\label{eq:ResidueRotation}
	R=\sin^{2}(\pi\omega_{0}) .
\end{equation}

An orbit that encircles an elliptic fixed point may have a rotation or winding number,
the average number of times the orbit goes around the fixed point per iteration.
A period-$q$ orbit always has rational winding number, $\omega = p/q$, where
$p \in \bZ$ represents the total rotation of the orbit per period. An
invariant circle enclosing the fixed point generically has an irrational winding number.
Indeed, Moser's twist theorem asserts that a fixed point of a smooth enough map  has a
neighborhood containing a Cantor set of invariant circles
when $\omega_0 \neq \tfrac12, \tfrac13$, or $\tfrac14$ and when a twist condition is satisfied
\cite{Moser1973,Meiss1992}.

An effective numerical technique for finding these invariant circles is based on Greene's residue criterion \cite{Greene1979}. This asserts that an invariant circle with irrational
winding number $\omega$ exists when periodic orbits in its neighborhood have bounded residue. In the case that an irrational torus is isolated---when it is locally robust---Greene found that the residues of periodic orbits that limit on the circle approach a limiting threshold value, $R \approx 0.25$. Conversely, he found that if the residues are unbounded, then there are no nearby invariant circles. Aubry-Mather theory implies that when the map has twist, each destroyed circle is replaced by a cantorus---topologically a Cantor set \cite{Meiss1992}.

Following the pioneering work of Greene, MacKay and Stark (GMS)  \cite{Greene1986}, we will use the residue criterion (see \Sec{Numerical}) to find ``boundary circles''; that is, circles that are locally isolated from one side. GMS showed that the rotation number $\omega_{BC}$ of a typical boundary circle is unusual from a number theoretic point of view; this can be seen most easily using the continued fraction expansion. Recall that any real number $\omega$
has a continued fraction expansion
\begin{equation}\label{eq:Continued Fraction}
	\omega=m_{0}+\frac{1}{m_{1}+\frac{1}{m_{2}+ \ldots}}
	      =[m_{0};m_{1},m_{2},\ldots]
\end{equation}
where $m_0 \in \bZ$ and $m_i \in \bN$ \cite{MR2445243,Khinchin1964}. When $\omega$ is irrational, this expansion is infinite, and a truncation after $n$ terms gives a rational approximation,
\begin{equation}\label{eq:convergents}
	\omega_{n}=\frac{p_{n}}{q_{n}}=[m_{0};m_{1}, m_{2},\ldots,m_{n}],
\end{equation}
called  the $n^{th}$ \emph{convergent} to $\omega$.
We will compute the rotation numbers for boundary circles for the H\'enon map and their continued fraction expansions in \Sec{BoundaryCircles}.

%%%%%%%%%%%%%%%%%
%% Henon Map
%%%%%%%%%%%%%%%%%
\subsection{H\'enon Map}

As a model, we will use H\'enon's famous area-preserving quadratic map \cite{Henon1969}; it is a one-parameter family that can be written in the form
\begin{equation}\label{eq:T}
	T:\left\{
	\begin{array}{ll}
		x'=& -y+2(K-x^{2})\\
		y'=&x
	\end{array}\right. .
\end{equation}
This map can be generated by the discrete Lagrangian
\[
	L(x,x')=xx'-2x(K-\frac{x^{2}}{3}),
\]
using the equations $y'=\frac{\partial L}{\partial x'}$ $y=-\frac{\partial L}{\partial x}$.

The map \Eq{T} has a  pair of fixed points when $K > -\tfrac14$. The point
\begin{equation}\label{eq:elliptic point}
	x^e = y^e \equiv \tfrac12 (-1+\sqrt{1+4K})
\end{equation}
has residue $R =\tfrac12 \sqrt{1+4K}$, and so
is elliptic when $-\tfrac14 < K < \tfrac34$, and the point
\begin{equation}\label{eq:hyperbolic point}
	x^h=y^h \equiv \tfrac12 (-1-\sqrt{1+4K})
\end{equation}
has residue $R = -\tfrac12\sqrt{1+4K}$, and so is hyperbolic.

The H\'enon map is reversible \cite{Devaney1984}, i.e., there exists an involution, $I$,
such that $T^{-1} = I \circ T \circ I$. Equivalently, $T$ can be factored into the product of two involutions, $T = (T\circ I) \circ I$ with
\begin{equation}\label{eq:involutions}
	\begin{split}	
	I:&\left\{ \begin{array}{ll}
			y'&=x\\
			x'&=y
		\end{array}\right. ,\\
	TI:&\left\{ \begin{array}{ll}
				y'&=y\\
				x'&=-x+2(K-y^{2})
		\end{array}\right. .
	\end{split}
\end{equation}
We will use these involutions to aid in finding periodic orbits, see \Sec{Numerical}.

%%%%%%%%%%%%%%%%%
%% The Meiss-Ott Model
%%%%%%%%%%%%%%%%%
\subsection{Islands Around Islands}\label{sec:Islands}

The elliptic fixed
point \Eq{elliptic point} is the ``parent" of an \emph{islands-around-islands} hierarchy:
it is typically encircled by other elliptic periodic orbits, which in turn are encircled
by yet more periodic orbits. Indeed, any elliptic period-$q$ orbit can be thought of
as a fixed point of $T^q$, and thus is also typically encircled by a family of
invariant circles that form a set of $q$-islands in the phase space.
The map $T^q$  also typically has periodic orbits, of period $q'$ say, that encircle the
original orbit. Thus this whole structure repeats on a smaller scale, giving $qq'$
islands-around-islands in the full phase space; an example is shown in \Fig{PhaseSpace7x7}.

%%%%%
\InsertFig{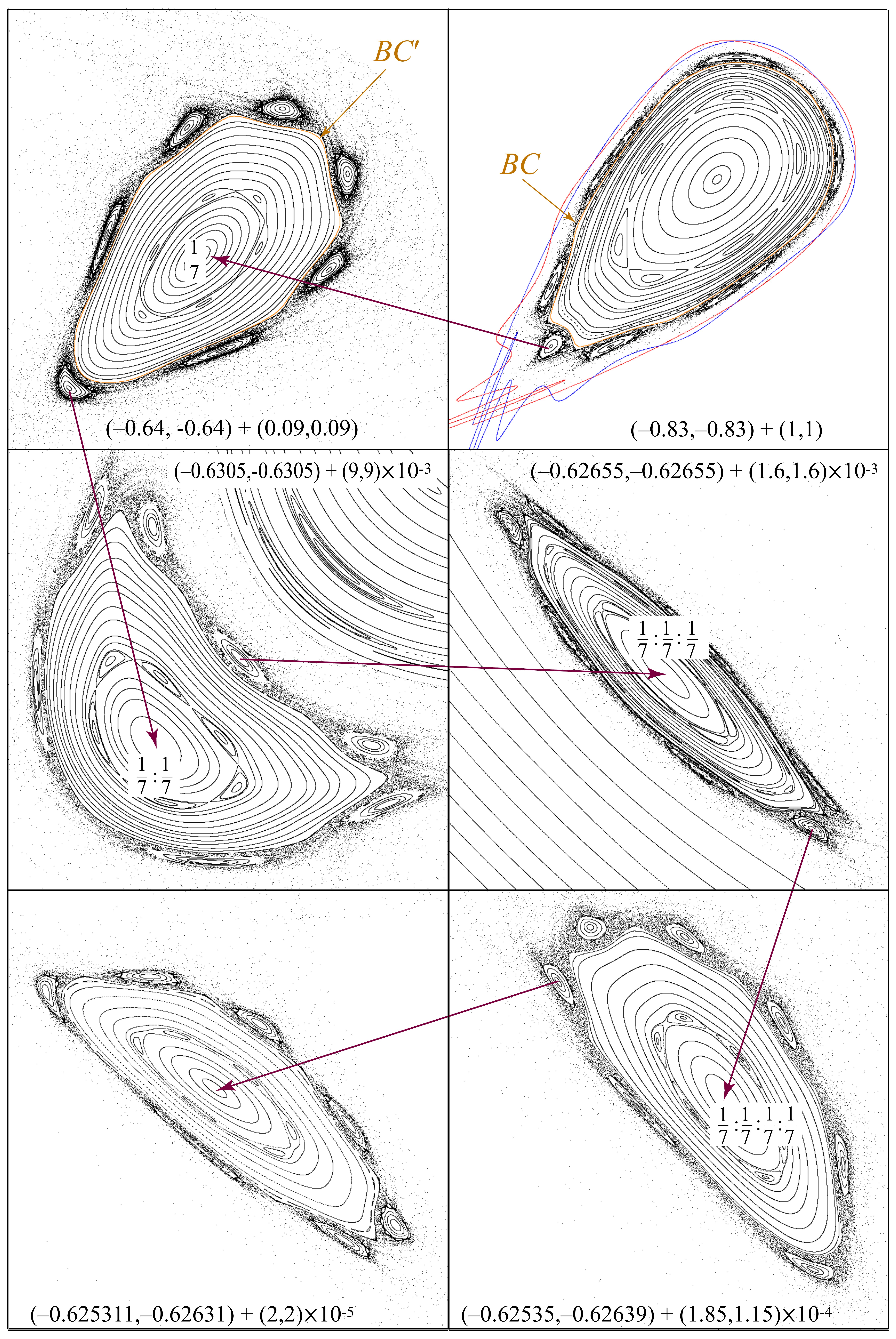}{(Color online) Six classes of the island around island hierarchy for \Eq{T} at $K=-0.17197997940$. The upper right panel shows the class-one island around $(x^e,y^e)$ and the stable and unstable manifolds of the hyperbolic fixed point. The state $S=1$ corresponds to a period-$7$ island chain (upper left); around each of these islands is period-$7$, class-two chain, the state $S= 10$. The class-one boundary circle, $BC$ is the labeled circle (brown) in the upper right panel, and the class-two boundary circle $BC'$ is in the upper left panel. Bounds of each panel are given in the form $(x_0,y_0) + (\Delta x, \Delta y)$ indicating the bottom left and width/height of the panel.} {PhaseSpace7x7}{0.7}
%%%%%

We are interested in how these islands are embedded in a connected chaotic component of the phase space. The boundary of a chaotic component consists of boundary circles: the outermost invariant circles surrounding elliptic periodic orbits. It seems to be generic that the boundary of a chaotic component consists of infinitely many circles \cite{Umberger1985}. Some of these are created by islands that encircle other islands---this is the case for \Fig{PhaseSpace7x7}, where each island has a set of period-$7$ islands outside its boundary circle.

This hierarchical structure was modeled as a tree by Meiss and Ott \cite{Meiss1986a}; the nodes of the tree correspond to a partition of a connected chaotic
component into regions that represent different stages of the island-around-island hierarchy,
see e.g., \Fig{SelfSimilarTree}.
This simplest version of this tree is binary; its nodes can be labeled by a binary sequence $S = s_0s_1\ldots$, with $s_i \in \{0,1\}$.
Each node represents a portion of the chaotic component surrounding a given periodic orbit. Following \cite{Meiss1986a}, the null state ``$\emptyset$'' denotes the root of the tree, it corresponds to the phase space outside the resonance zone formed by the stable and unstable manifolds of the hyperbolic fixed point (the thin red and blue curves in \Fig{PhaseSpace7x7}). Orbits in this region can be unbounded, and we view $\emptyset$ as absorbing: whenever an orbit leaves the resonance zone, iteration stops.

%%%%%%%
\InsertFig{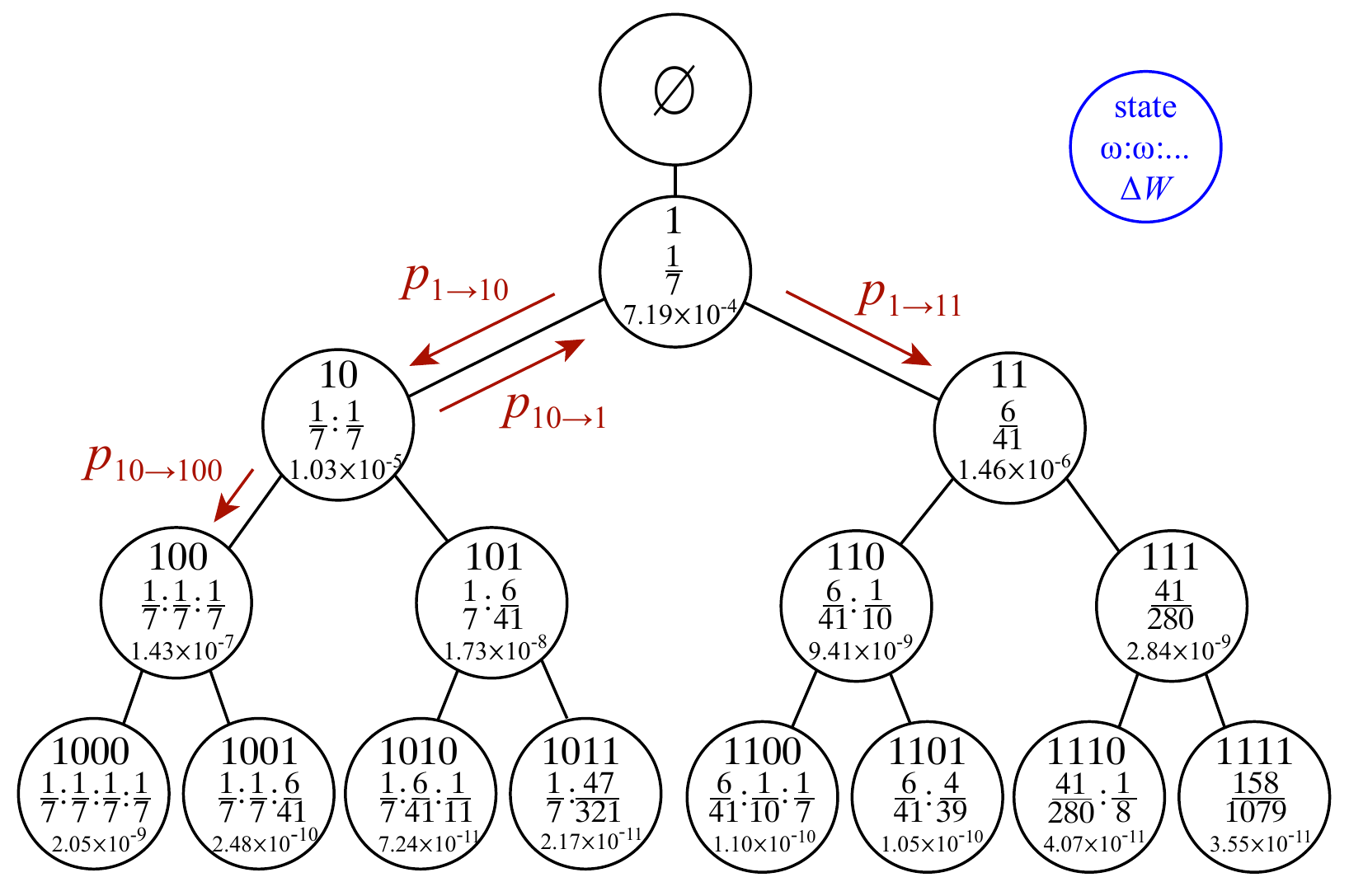}{(Color online) Tree for the H\'enon map at $K=-0.17197997940$ with the
phase space shown in \Fig{PhaseSpace7x7}.
Each node is labeled by the state $S$, the winding number sequence for each class, $\frac{p_1}{q_1}: \frac{p_2}{q_2} : \ldots$ and the flux \Eq{Flux} through the pair of periodic orbits. Several transition probabilities, $p_{S,S'}$, are also indicated. The class one boundary circle for this case has rotation number
$\omega_{BC} = [0;6,1,4,1,5,1,2,1,\ldots]$.}{SelfSimilarTree}{0.8}
%%%%%%%

The boundary circle of the elliptic fixed point corresponds to the largest invariant circle that encloses the fixed point, we denote it by $BC$.
The states below $\emptyset$ on the tree represent a partition of the chaotic component outside $BC$ and inside the main resonance zone. Some of periodic orbits that rotate about the elliptic fixed point are outside $BC$. Indeed, since the rotation number is typically a monotone decreasing function of distance, the rotation numbers of the orbits encircling the fixed point typically decrease with distance from the elliptic point, converging to zero as they approach the broken separatrix, see \Fig{OmegaMonotone}.\footnote
%%%%%
{Note that $\omega$ does not always decrease monotonically; in particular, this does not occur near twistless bifurcations \cite{Moeckel1990,Dullin2000}.}
%%%%%
Between every pair of rational rotation numbers, there are infinitely many irrationals and these correspond to cantori that encircle $BC$. The flux through these cantori is a rapidly varying function of their rotation number; it typically has sequence of sharp local minima as $\omega \to \omega_{BC}^{-}$ \cite{MacKay1984, Meiss1986a}. The states correspond to chaotic regions around rational rotation numbers that are bounded by these locally minimal-flux cantori. This gives a sequence of ``levels,'' states that limit on a given boundary circle.

%%%%
\InsertFig{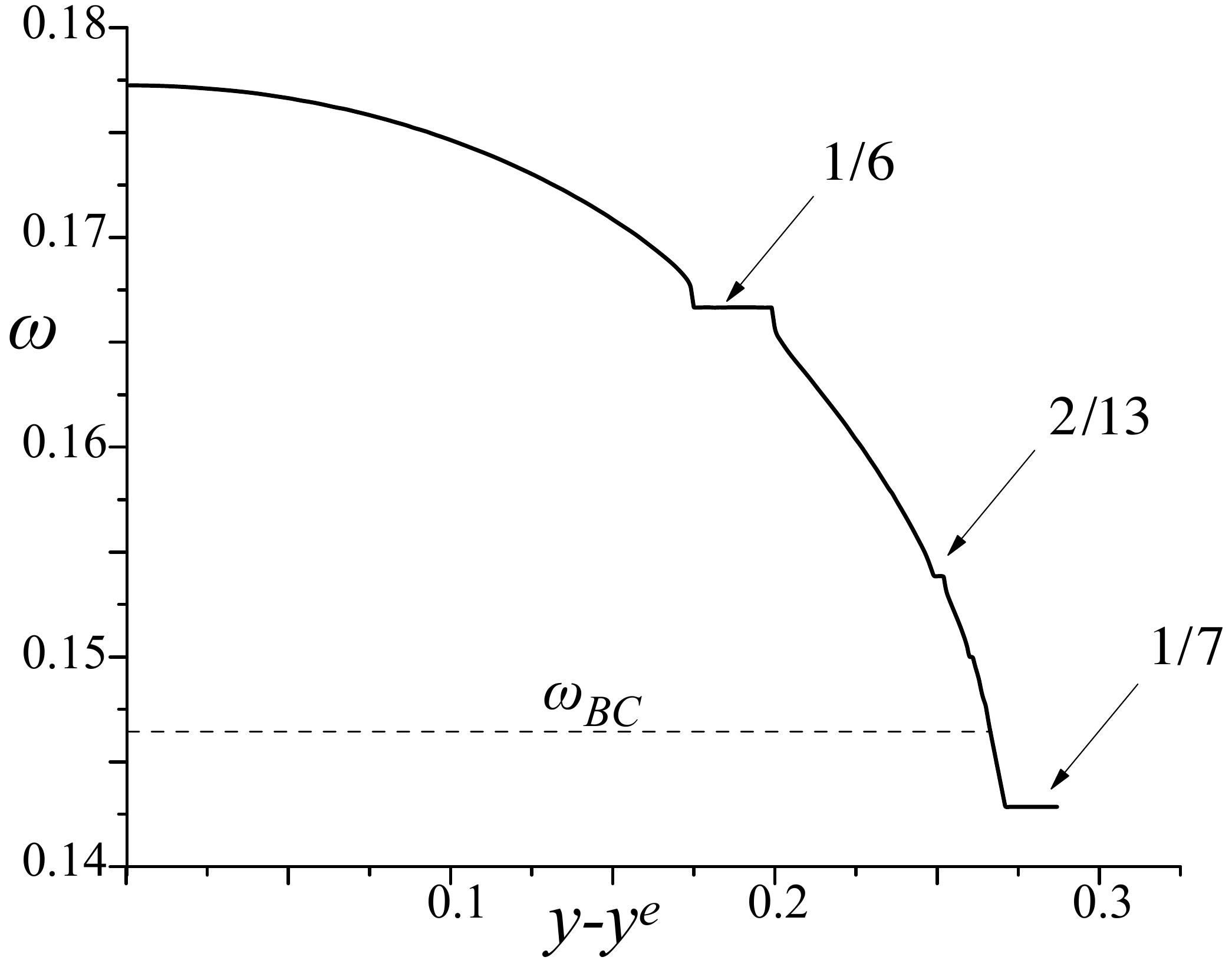}{The winding number $\omega$ as a function of the distance from the elliptic point for \Eq{T} at $K=-0.17197997940$. The rotation number decreases monotonically, with flat spots marking the location of islands.}{OmegaMonotone}{0.6}
%%%%

One way to choose the states is to use the periodic orbits with rotation numbers that are the convergents \Eq{convergents} of the continued fraction expansion of $\omega_{BC}$.
These rational rotation numbers are alternately larger than and smaller than $\omega_{BC}$ \cite[Thm. 4]{Khinchin1964}. Indeed, \Eq{Continued Fraction} implies that $\omega_{1}=m_0 + \tfrac{1}{m_1}$ and $\omega_{2}=m_0 + \tfrac{m_2}{m_1m_2+1}$ so that  $\omega_{1}>\omega_{BC} > \omega_{2}$. More generally one can see that the convergents in the outer region, where $\omega < \omega_{BC}$, correspond to even $i$. For example, in \Fig{PhaseSpace7x7}, $\omega_{BC} = [0;6,1,4,1,5,1,2,1,\ldots]$. The first convergent $\tfrac{p_1}{q_1} = \tfrac16 = [0;6]$, gives rise to a period-$6$ island chain in the trapped region, while the second $\tfrac17 = [0;6,1]$ gives a chain embedded in the outer, chaotic component. These correspond to the largest steps seen in \Fig{OmegaMonotone}. Subsequent even convergents, e.g., $\tfrac{6}{41}=[0;6,1,4,1]$, $\frac{41}{280}$, $\frac{158}{1079}$, etc.,  result in smaller islands closer to the boundary circle that are too small to be seen in the figure.
For the tree representation, incrementing the level corresponds to a transition to the right, and will be denoted with the symbol ``$1$''. This sequence of convergents gives the first branch of the tree, as shown in \Fig{building_the_tree}; each of these states belongs to class one. The first four of these states are indicated in \Fig{SelfSimilarTree}.

When the positive residue periodic orbit defining a level is elliptic, it is surrounded by an island of ``class two". For example if the state $S=1$ with rotation number $p/q$ is elliptic, then the map $T^q$ has $q$-fixed points surrounded by islands. The boundary circle about this family will have some rotation number, $\omega_{BC'}$, say, relative to $T^q$. The outer convergents to $\omega_{BC'}$ correspond to states of ``class two,'' and to a step to the left on the binary tree (denoted by the symbol ``$0$''). Thus the period-$7$ island around island the first period-$7$ island is the $\tfrac17: \tfrac17$ orbit; it corresponds to a state of $49$ islands for $T$ and to the state $S=10$ on the tree. Subsequent outer convergents to the boundary circle of the period-$7$ island correspond to the states $S = 101^\ell$, $\ell=1,2,\ldots$, as shown on the right panel of \Fig{building_the_tree}.

In general a state $S =s_1 s_2 s_3, \ldots$ has a class equal to one plus the number of zeros  in $S$, and a level equal to the number of ones in $S$. For example, the state $S=1011$ corresponds to a periodic orbit of class two and level three.

%%%%%
\InsertFig{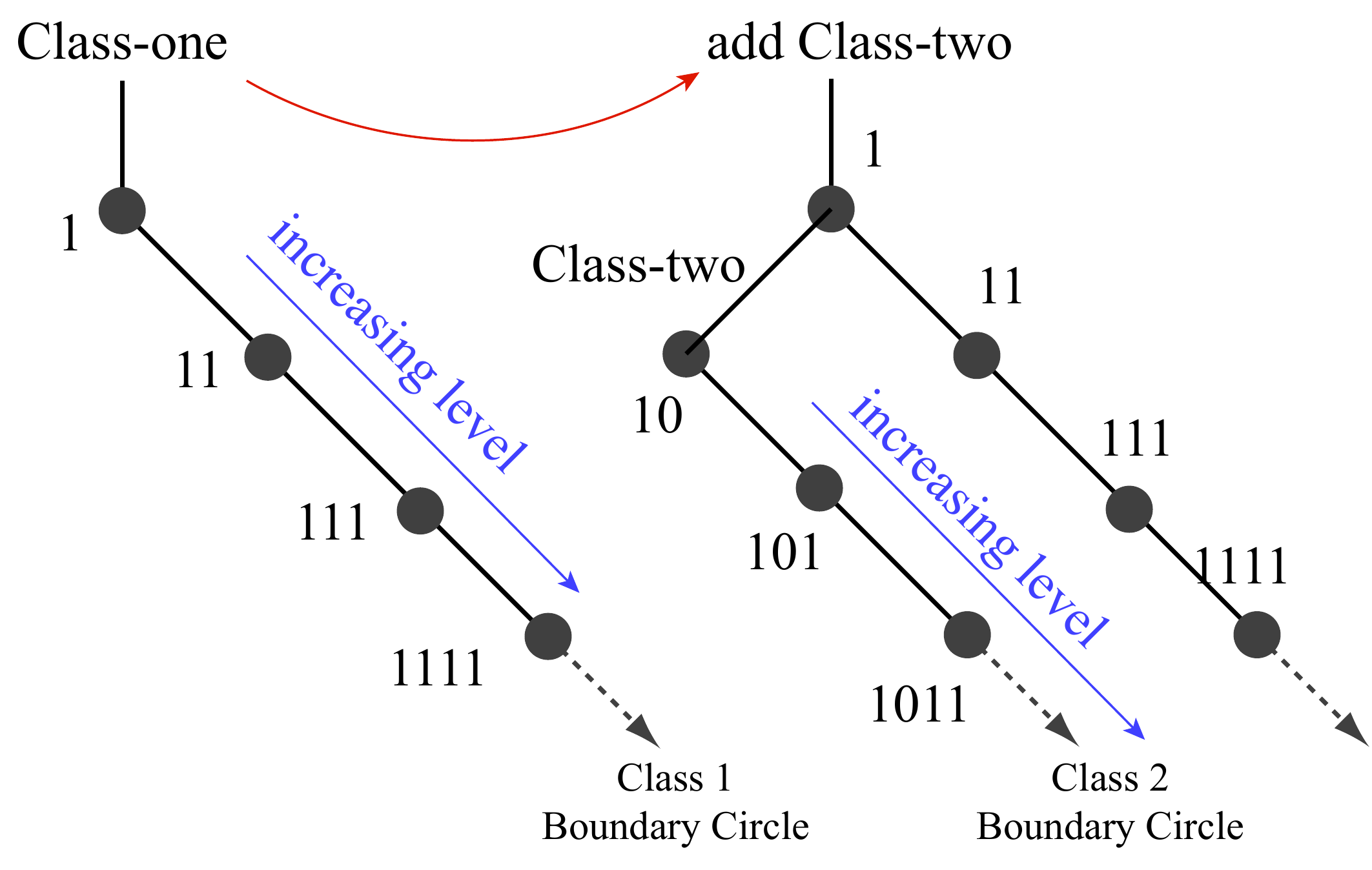}{(Color online) Schematic diagram depicting the binary level-class tree.
Outer convergents to the class-one boundary circle give the rightmost
branch of the tree, the states $S=1^j$. When there is an island in this component, it gives rise to states of class-two with first approximant $S=10$, and successive convergents giving states $101^j$
that converge to the class-two boundary circle.}
{building_the_tree}{0.6}
%%%%%

%%%%%%%%%%%%%%%%%
%% Meiss Ott Model
%%%%%%%%%%%%%%%%%
\subsection{Markov Tree Model}\label{sec:MeissOtt}
To study transport in a connected chaotic component with an infinite set of islands-around-islands, Meiss and Ott introduced a Markov model based on the tree structure discussed above
\cite{Meiss1986a}. Here we recall some of the notation for this model.

A Markov model for transport on the tree is defined by transition probabilities
$p_{S\to S'}$ for each pair of connected nodes on the tree, recall \Fig{SelfSimilarTree}. The probability of such a transition is determined by the flux of trajectories that move from one state to another, and this limited by the area of the turnstile \cite{MacKay1984} in the most resistant cantorus that divides the states. We denote this flux by $\Delta W_{S,S'} = \Delta W_{S',S}$; it is symmetric because the net flux through any region of phase space must be zero. The flux through a cantorus can be computed by the MacKay-Meiss-Percival action principle \cite{MacKay1984}, see \Sec{Numerical}. It is known that the average transit time through a state bounded by such partial barriers exactly equal to  the area $A_S$ of the accessible region of phase space in the state $S$ divided by the exiting flux \cite{Meiss1997}. The Markov approximation is to assume that these  transit times are long enough that correlations are unimportant, and so that the transition probability is
\[
	p_{S,S'} = \frac{\Delta W_{S,S'}}{A_S}
\]

The only nodes that are connected on the tree
are parent-daughter nodes. The daughters of a state $S = s_1s_2s_3 \ldots s_j$ are denoted by concatenation: $S0$ and $S1$. The unique parent of $S$, obtained by deleting the last symbol, is denoted $DS$.
There are two important transition probabilities, $p_{S\to DS}$ for moving ``up'' from state $S$ to its parent, and $p_{DS \to S}$ for moving ``down'' from parent to daughter.
It is convenient to categorize the change in transition probabilities from state to state by the two ratios
\begin{align}
	w_{S}^{(i)}&= \frac{p_{S \to Si}}{p_{S \to DS}} = \frac{\Delta W_{S,Si}}{\Delta W_{S,DS}} ,\label{eq:FluxRatio}\\
	a_{S}^{(i)}&= \frac{p_{S\to Si}}{p_{Si\to S}} = \frac{A_{Si}}{A_{S}}  . \label{eq:AreaRatio}
\end{align}
These ratios are measures of the asymmetry between motion ``up'' and ``down'' the tree.

Meiss and Ott, assumed that the tree is self-similar. In this case the ratios \Eq{FluxRatio}-\Eq{AreaRatio} are independent of the state $S$, though they depend on the choice of class, $i=0$, or level, $i = 1$. Self-similarity only occurs for special cases. The renormalization theory of MacKay and Greene implies that for a critical noble invariant circle, the level hierarchy is asymptotically self-similar \cite{MacKay1983}. Similarly, just as for the Feigenbaum period-doubling scenario, there are cases in which the class hierarchy will be asymptotically self-similar \cite{Meiss1986}. The case of a self-similar tree with seven islands as the outermost convergent of each boundary circle was shown in \Fig{PhaseSpace7x7}. Note, however, that for a generic choice of parameter value, neither self-similar scenario will hold.
An example of this more typical behavior is shown in \Fig{TreeK01995}. Here, unlike in \Fig{PhaseSpace7x7}, the number of islands and the rotation number of the boundary circles vary with class.

%%%%%
\InsertFigTwo{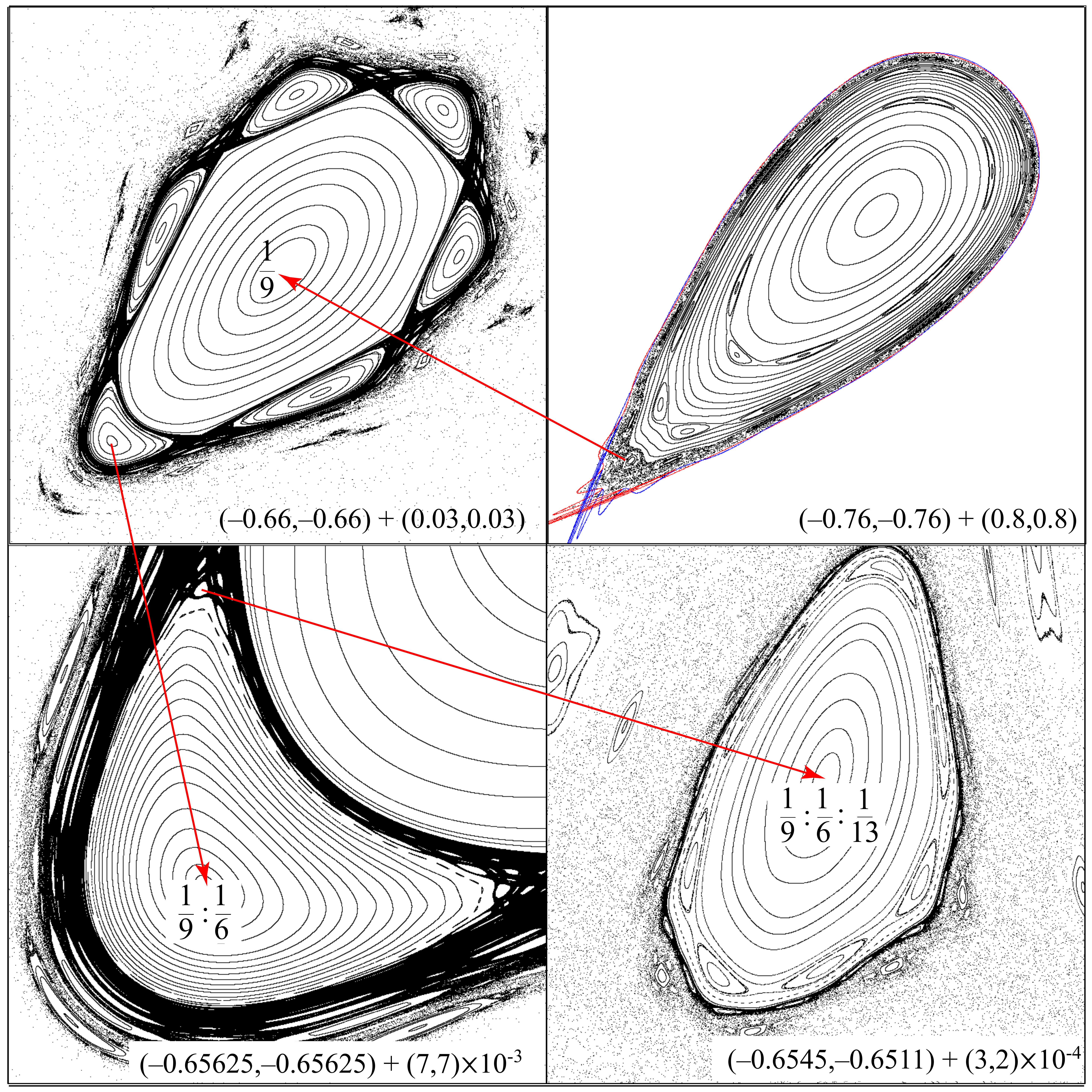}{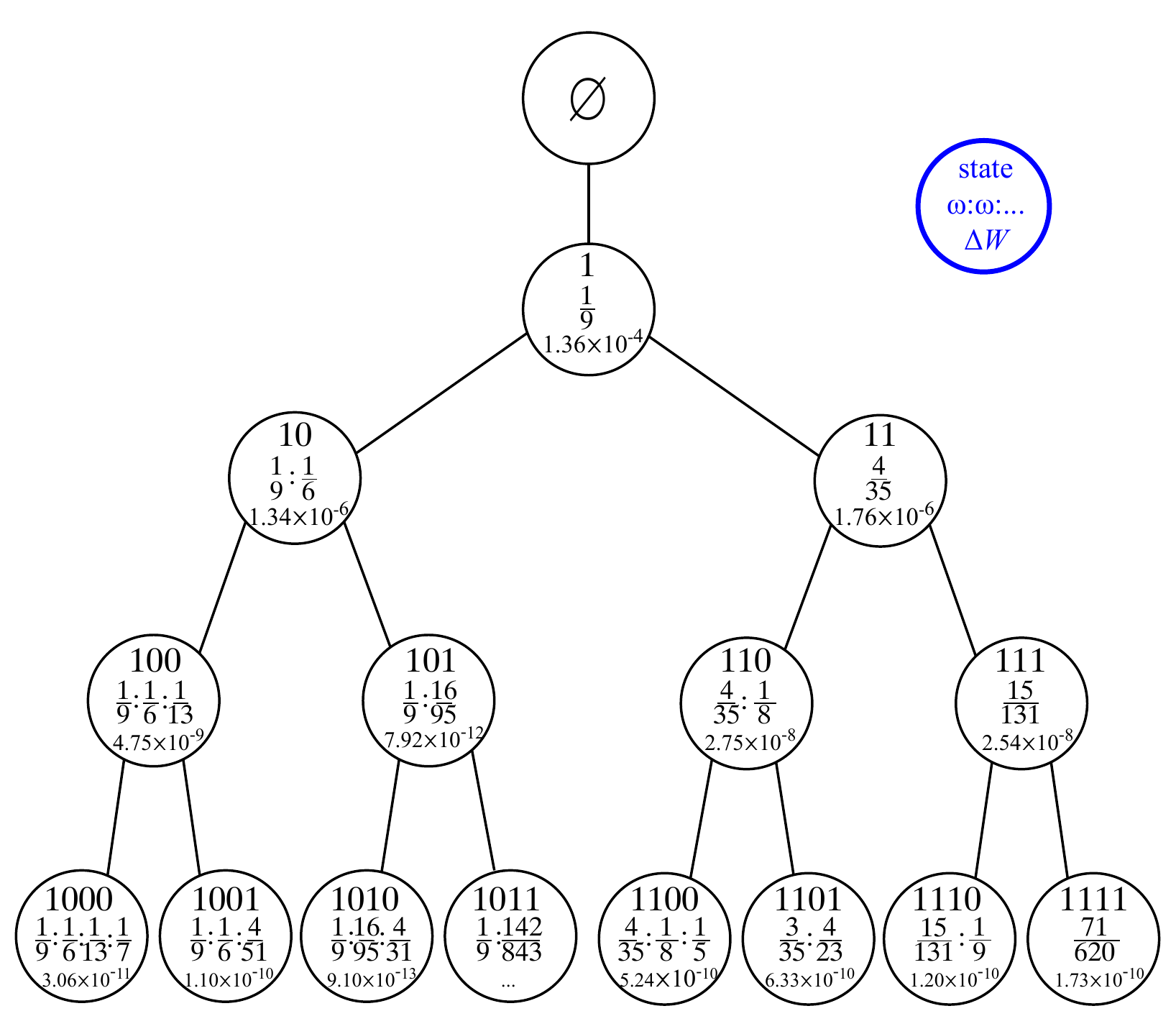}{(Color online) Four classes of the islands-around-islands
hierarchy for \Eq{T} with $K = -0.1995$, and the corresponding tree. Unlike Figs.~\ref{fig:PhaseSpace7x7}-\ref{fig:SelfSimilarTree}, the tree is not self-similar.
Here the class-one boundary circle has rotation number $\omega_{BC} = [0; 8, 1, 2, 1, 2, 1, 3,1,\ldots]$.}{TreeK01995}{0.45}
%%%%%%%%%%%%%%%%

The fluxes $\Delta W_{S,S'}$ are given by the difference between the action
of the minimal flux cantorus between $S$ and $S'$, and
the action of its homoclinic, minimax orbit \cite{MacKay1984}.
Since the cantori are difficult to compute, we will compute instead the
flux through the periodic orbits themselves. If $\{\vec{x}_t^e: t = 0,\ldots, q_S-1\}$ and $\{\vec{x}_{t}^{h}: t = 0,\ldots,q_S-1\}$ denote the
elliptic and hyperbolic period $q_S$ orbits for state $S$, then
\begin{equation}\label{eq:Flux}
	\Delta W_S =\sum_{t=0}^{q_S-1}\left[L(\vec{x}_{t}^{e})-L(\vec{x}_{t}^{h})\right]
\end{equation}
is the flux through the periodic orbit pair.
We replace \Eq{FluxRatio} with
\begin{equation}\label{eq:PeriodicFluxRatio}
	w_S^{(i)} = \frac {\Delta W_{Si}}{\Delta W_S},
\end{equation}
the rate of the decrease in flux through a daughter periodic orbit relative to its parent.

If the tree were self-similar, the ratios \Eq{FluxRatio} and \Eq{PeriodicFluxRatio} would be the same---everything is geometrically similar under the
renormalization scaling coefficients \cite{Meiss1986a}. When the tree is not self similar,
this is no longer the case. Therefore the flux ratios found in this paper are only an approximation of those that dictate the dynamics on the Markov tree.

We will investigate in \Sec{Flux} the variation of the ratio \Eq{PeriodicFluxRatio} with the state $S$.

%%%%%%%%%%%%%%%%%
%% Numerical Methods
%%%%%%%%%%%%%%%%%
\section{Numerical Methods}\label{sec:Numerical}

%%%%%%%%%%%%%%%%%
%% Finding POs
%%%%%%%%%%%%%%%%%
\subsection{Finding periodic orbits}\label{sec:FindingPOs}

Periodic orbits can be most easily found using the symmetries of the
reversible map as described in \cite{Meiss1986}. The fixed sets of the involutions \Eq{involutions} define four rays emanating from the elliptic fixed point  \Eq{elliptic point}:
\begin{equation}\label{eq:symmetry lines}
\begin{split}
	\Fix{I}^{+}  &= \{(y,y): y > y^e\}\\
	\Fix{I}^{-} &= \{(y,y): y < y^e\}\\
	\Fix{TI}^{+} &= \{(K-y^{2},y): y > y^e\}\\
	\Fix{TI}^{-}&= \{(K-y^{2},y): y < y^e\}\\
\end{split} .
\end{equation}
Note that $\Fix{I}^{+}$ and $\Fix{I}^{-}$ are together comprise
the fixed set of the involution $I$, and correspondingly for $TI$. It was found in \cite{Meiss1986} that all
positive residue, class-one periodic orbits have a point on the dominant symmetry set $\Fix{TI}^{+}$. Hence
if $(x_q(y),y_q(y)) = T^q(K-y^2,y)$, one must solve the single equation
\begin{equation}\label{eq:periodic orbit finding}
	y_q(y) - y  = 0
\end{equation}
for $y$, the position along the symmetry line.

A solution of \Eq{periodic orbit finding} is found iteratively using a secant method
with a trust region. As the first guess we use the distance between the parent
elliptic point and the corresponding hyperbolic point,
scaling it by an ad hoc power law:
\begin{equation}\label{eq:guess}
	y=y^e \pm (y^h-y^e)(1-\tfrac{\omega}{\omega_{0}})^{a} .
\end{equation}
where $\omega$ is the winding number of the desired orbit around its parent, $\omega_{0}$ is the linearized winding number of parent, and we chose $a =10$. The sign in \Eq{guess} is determined according to the location on the symmetry line of the daughter periodic orbit relative to the parent periodic orbit point which lies on the same symmetry line. The symmetry line on which the daughter elliptic periodic orbit is found determined as in \cite{Meiss1986}.
%%%%
This equation should be reasonable for the case of an island with a rotation number that monotonically decreases to zero as the hyperbolic point is approached.

The trust region for the secant method is determined using the rotation number
of the orbit, computed using polar coordinates
about its parent elliptic point. That is, an orbit with rotation number $\omega = p/q$ must rotate $p$ times around the parent elliptic point in $q$ iterations. The first two convergent orbits on the opposite sides of
a boundary circle (see \Sec{Boundary}) define the
trust region, assuming that the rotation number is indeed monotonic.
If this method fails, we use simulated annealing to try additional guesses.

To find an orbit of higher class, one has to provide a guess, \Eq{guess}, based on the location of the parent elliptic and hyperbolic points. For that one has to choose one point from each of these orbits that are in the same ``island." This is done by simply choosing the iterates of the two orbits that minimize the angle between them in polar coordinates, relative to their parent elliptic point.

%%%%%
%%%%%%%%%%%%%%%%%
%% Boundary Circles
%%%%%%%%%%%%%%%%%
\subsection{Finding Boundary Circles}\label{sec:Boundary}

We compute  boundary circles using the method of GMS \cite{Greene1986, Meiss1986}.
Given a pair of positive residue periodic orbits of the same class (encircling a given parent) with rotation numbers $\tfrac{p}{q}$ and $\tfrac{p'}{q'}$
that are Farey neighbors ($pq'-qp' = \pm1$), and $q'>q$, their mean residue is defined by
\begin{equation}\label{eq:meanResidue}
		R^* = (R R'^\phi)^{\phi^{-2}} ,	
%	\log(R^{*})=\frac{\log(R)+\phi\log(R')}{\phi^{2}},
\end{equation}
where $\phi$ is the golden mean, so that $\phi^2 = \phi +1$. GMS found that, asymptotically, there is an invariant circle between the neighbors if $R^*$ is much smaller than $0.25$. We will use the threshold $R_{th} = 0.3$, which gives faster convergence to the boundary circle, recall \cite{Meiss1986}.

Assuming that the rotation number is a monotone decreasing function of distance from the parent orbit, the goal is to find the invariant circle with the smallest rotation number.
If the two orbits have small enough mean residue, then the interval $[\tfrac{p}{q},\tfrac{p'}{q'}]$ may contain an invariant circle. To test this, we find the Farey daughter orbit, with rotation number  $\frac{p''}{q''}=\frac{p+p'}{q+q'}$ and then compute mean residue of the two sub-intervals. If the lower rotation number interval has $R^* <R_{th}$, we take that interval as a candidate containing the boundary, and divide again. If not, we take the higher interval. If both fail, we go one step back and take an interval with larger rotation number. This gives the Farey sequence of a candidate boundary circle which is closely related to its continued fraction expansion, recall \cite{Greene1986}.

%%%%%%%%%%%%%%%%%
%% Building the Tree
%%%%%%%%%%%%%%%%%

\subsection{Building The Markov Tree}\label{sec:Building}

Using the method described above we found periodic orbits that approximate the class-one boundary circle for \Eq{T} for a uniform grid of $4000$ values of $K \in [-0.25, 0.75]$. Orbits with periods up to $10^{5}$ were found, giving continued fraction expansions with $8$ to $18$ coefficients.  Recall that the convergents to the class-one boundary circle describe the rightmost branch of the tree (states $S= 1^j$ in \Fig{building_the_tree}).
If the positive residue orbit for a state $S$ is elliptic ($0<R<1$),
it is encircled by an island,
and we next find the convergents that approximate its class-two boundary circle.
These correspond to the states $S0, S01, S011, \ldots$. We found boundary
circles up to class four.

For the $4000$ values of $K$, the algorithm found $2667$ class-one boundary circles, their rotation numbers are shown as a function of $K$ in \Fig{Omega1VsK}. The algorithm fails to find periodic orbits in two intervals of $K$, the largest being $(0.1920,0.4033)$. The failure is due to twistless bifurcations in
the neighborhood of tripling ($K= \tfrac{5}{16}$) and quadrupling ($K = 0$) points \cite{Dullin2000,Moeckel1990} that cause the rotation number to not be a monotone function.
The figure also shows the rotation number $\omega_2 = \frac{p_2}{q_2}$ of the outermost convergent (black squares and red/light-gray circles). There are  are ranges of $K$ for which the positive residue orbit in this state is unstable, indicated by the red/light-gray points in the figure. This especially occurs near the tripling bifurcations of the elliptic fixed point. When the state $S=1$ is not elliptic, it has no class-two island, and so the state $S=10$ and all of its descendants do not exist.  An example of such a tree is shown in \Fig{TreeK0575}.

One implication of this is that the number of computed class-two boundary circles is reduced: the algorithm found $5$-$16$ convergents for $1290$ class-two boundary circles, the states $101^j$.

%%%%%
\InsertFig{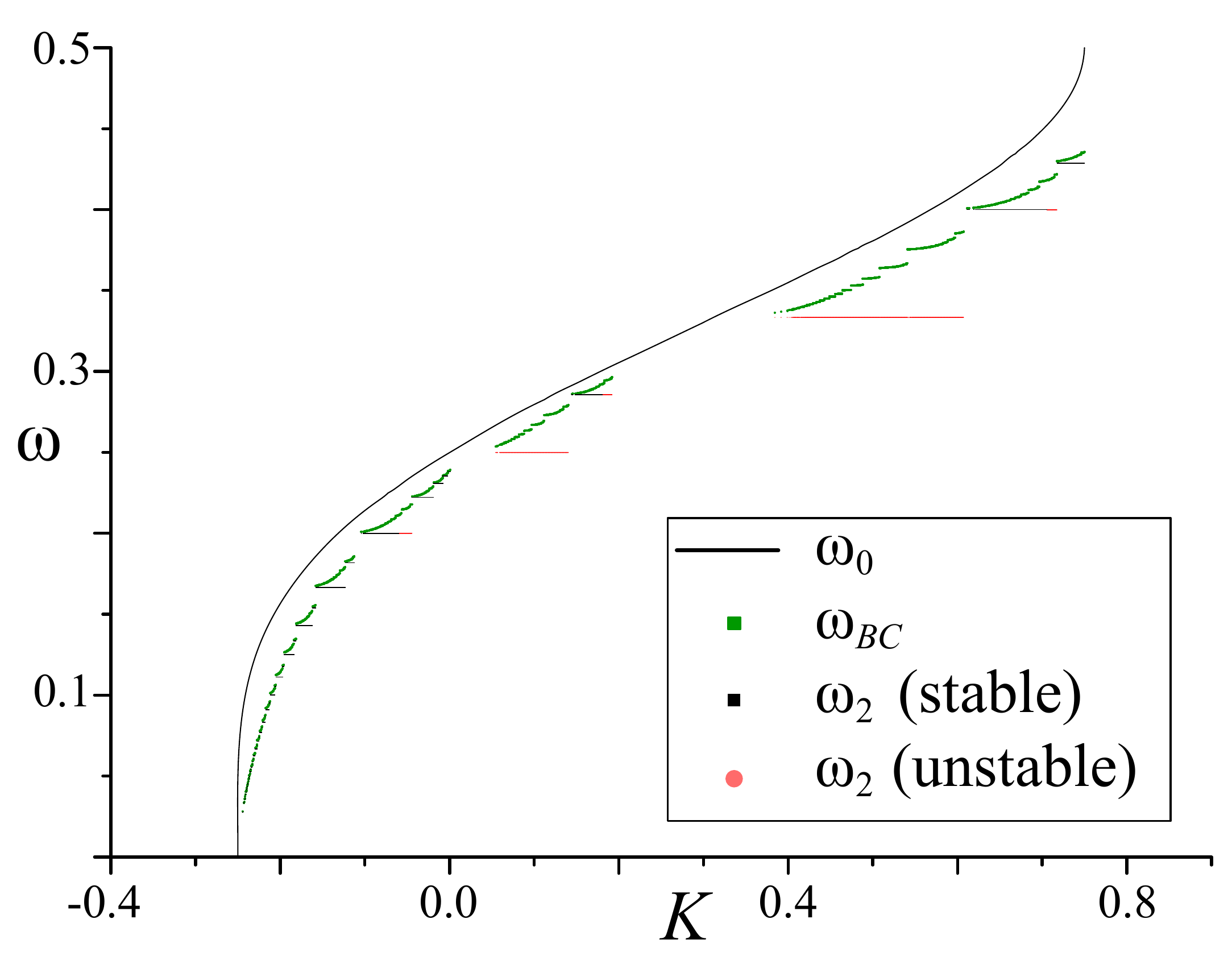}{(Color online) Rotation numbers for the H\'enon map \Eq{T} as a function of $K$.  The solid curve shows the rotation number $\omega_0$ of the elliptic fixed point, \Eq{ResidueRotation}, and the dark-gray (green) squares show the rotation number $\omega_{BC}$ of the class-one boundary circle. The lowest points show the rotation number $\omega_2$ of outermost convergents to the boundary circle;  black squares denote stable ($0<R<1$) and light-gray circles (red), unstable ($R>1$) orbits. When the orbit is unstable, the state $10$ and all of its descendants do not exist.}{Omega1VsK}{0.8}
%%%%%

%%%%%%%%%%%%%%%%
\InsertFigTwo{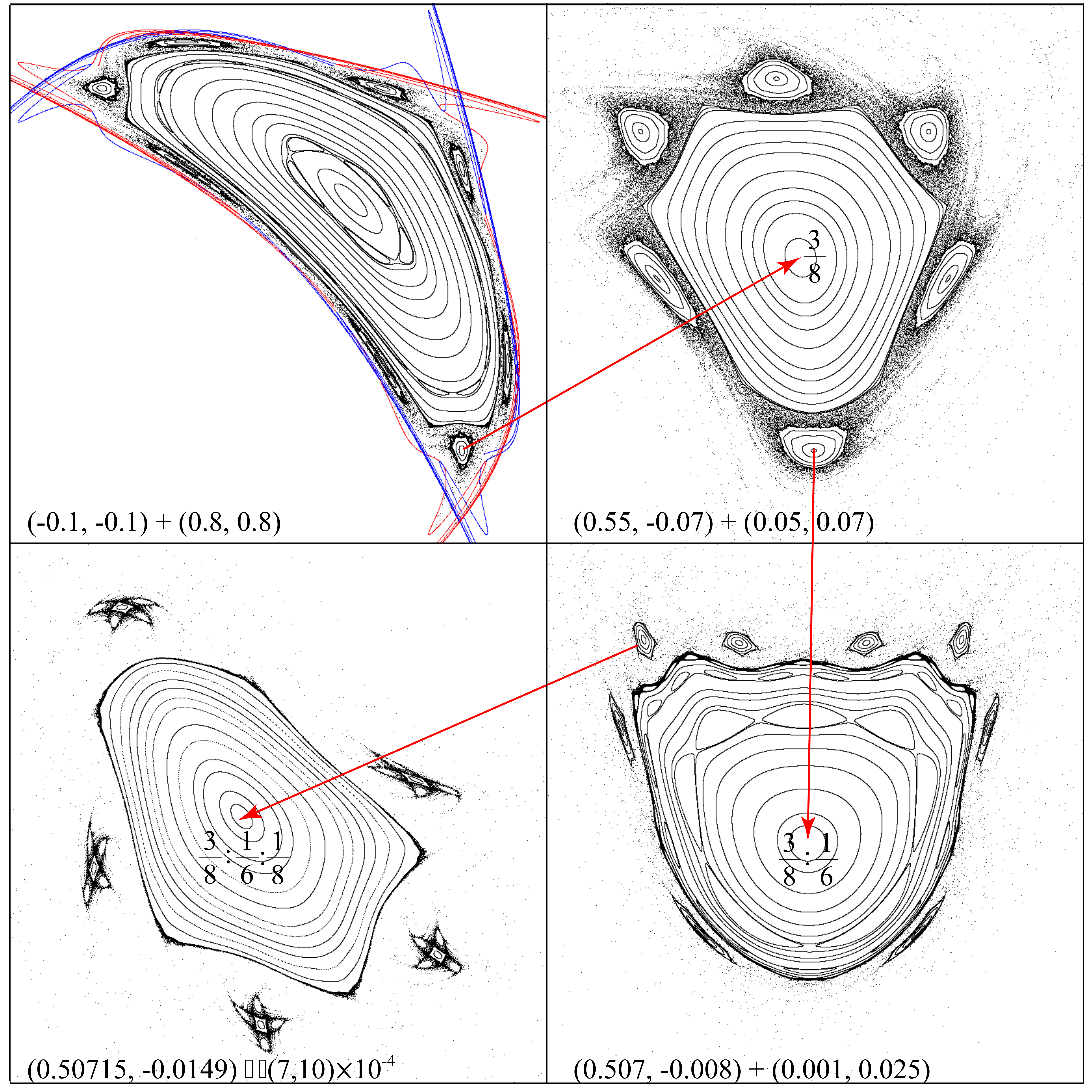}{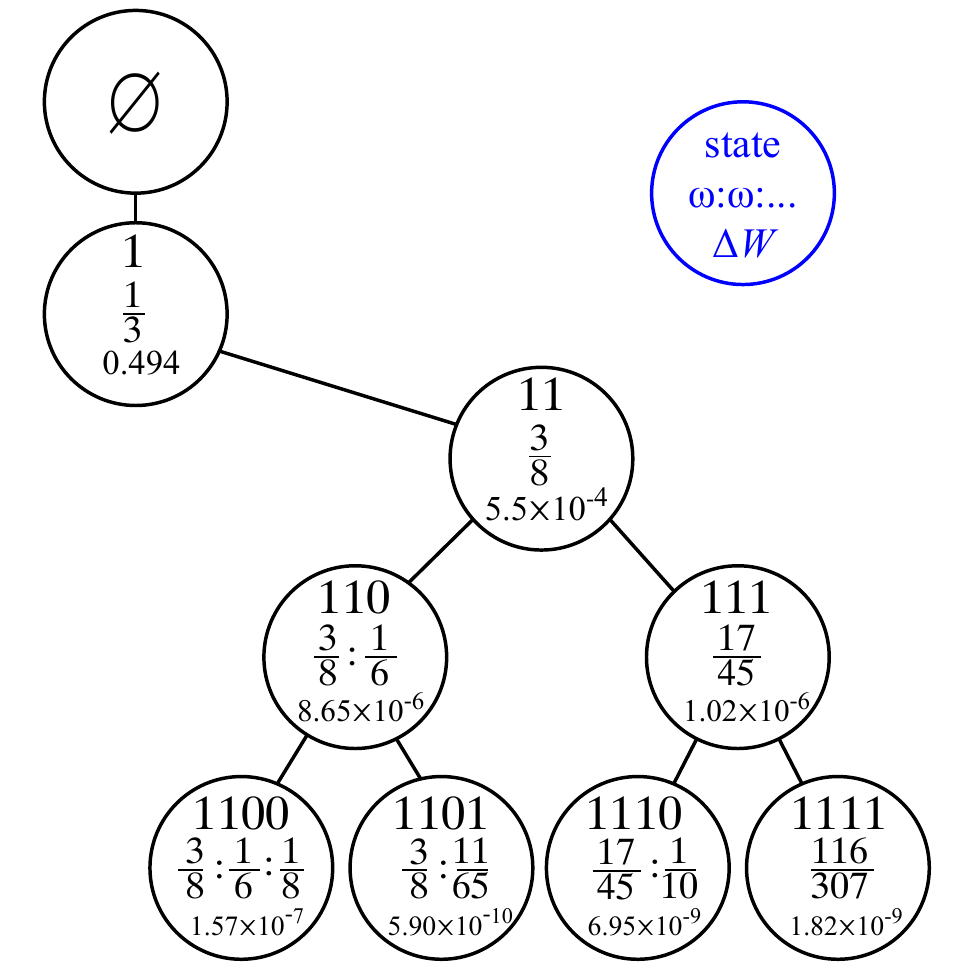}{(Color online) 
Island hierarchy for \Eq{T} when $K = 0.575$.
The tree is incomplete because the positive residue periodic orbit, with rotation number $\tfrac{1}{3}$ at state $S=10$ is unstable. Nevertheless, the $11$ state has a period-$8$ island, with a corresponding class hierarchy. The class-one boundary circle has rotation number $\omega_{BC} = [0;2,1,1,1,4,1,5,1,\ldots]$}{TreeK0575}{0.45}
%%%%%%%%%%%%%%%%

%%%%%%%%%%%%%%%%%
%% Data
%%%%%%%%%%%%%%%%%
\subsection{Error Analysis}\label{sec:ErrorAnal}

Once a tree---the corresponding periodic orbits, residues, and fluxes---was computed, it was found that for some states the ratio \Eq{PeriodicFluxRatio} was larger than one. Since increasing level or class should result in smaller areas, such a ratio signals an anomaly. By investigating some special cases, we found that this was a numerical error and there were two possible problems. The first is that $\Delta W_S$ can not be computed if it is too small relative to the accumulated numerical error of the orbit finding routine. To deal with this, we discarded all data from orbits for which
\[
	\Delta W_S <200\epsilon q_S ,
\]
where $q_S$ is the period of the orbit and $\epsilon = 10^{-15}$ is the error bound of the secant method solver.
The second problem is that the last few continued fraction coefficients of the rotation number of the boundary circle  may be incorrect due to the arbitrary choice $R_{th} = 0.3$ of the threshold residue. To deal with this, we discarded the last two coefficients computed. These corrections removed most of the anomalous values.

%%%%%%%%%%%%%%%%%
%% Boundary Circles
%%%%%%%%%%%%%%%%%
\section{Rotation Numbers of Boundary Circles}\label{sec:BoundaryCircles}

In this section we study the distribution of the continued fraction
coefficients $m_{i}$ for the rotation numbers of the boundary circles
of the H\'enon map \Eq{T}. The results are compared to the distribution that
would occur if the rotation numbers were real numbers
selected at random with uniform measure, namely, the Gauss-Kuzmin (GK) distribution \cite{Khinchin1964},
\begin{equation}\label{eq:Gauss-Kuzmin}
	p_{GK}(n)=p_{GK}(m_{i}=n)=-\log_{2}\left(1-\frac{1}{(1+n)^{2}}\right) .
\end{equation}

In their pioneering work, Greene, MacKay and Stark (GMS) \cite{Greene1986} computed some $400$ boundary circles, mostly for Chirikov's standard map with $K \in\left[0.75,0.97\right]$. They found $5$-$12$ elements for the rotation number of each boundary circle and computed separate distributions for the outer (even) and inner (odd) elements. They conjectured that for the outer (even) coefficients only the elements $m_{i} =1$ and $2$ occur, while for the inner (odd) coefficients only elements $m_i \le 5$ occur. Note that these observations contrast with \Eq{Gauss-Kuzmin} where the probability of arbitrarily large elements is nonzero. Making use of the enormous improvement in computing power since 1986, when \cite{Greene1986} was published, we explore here the distribution of $m_{i}$ for the H\'enon map \Eq{T}.

%%%%%
\InsertFigTwo{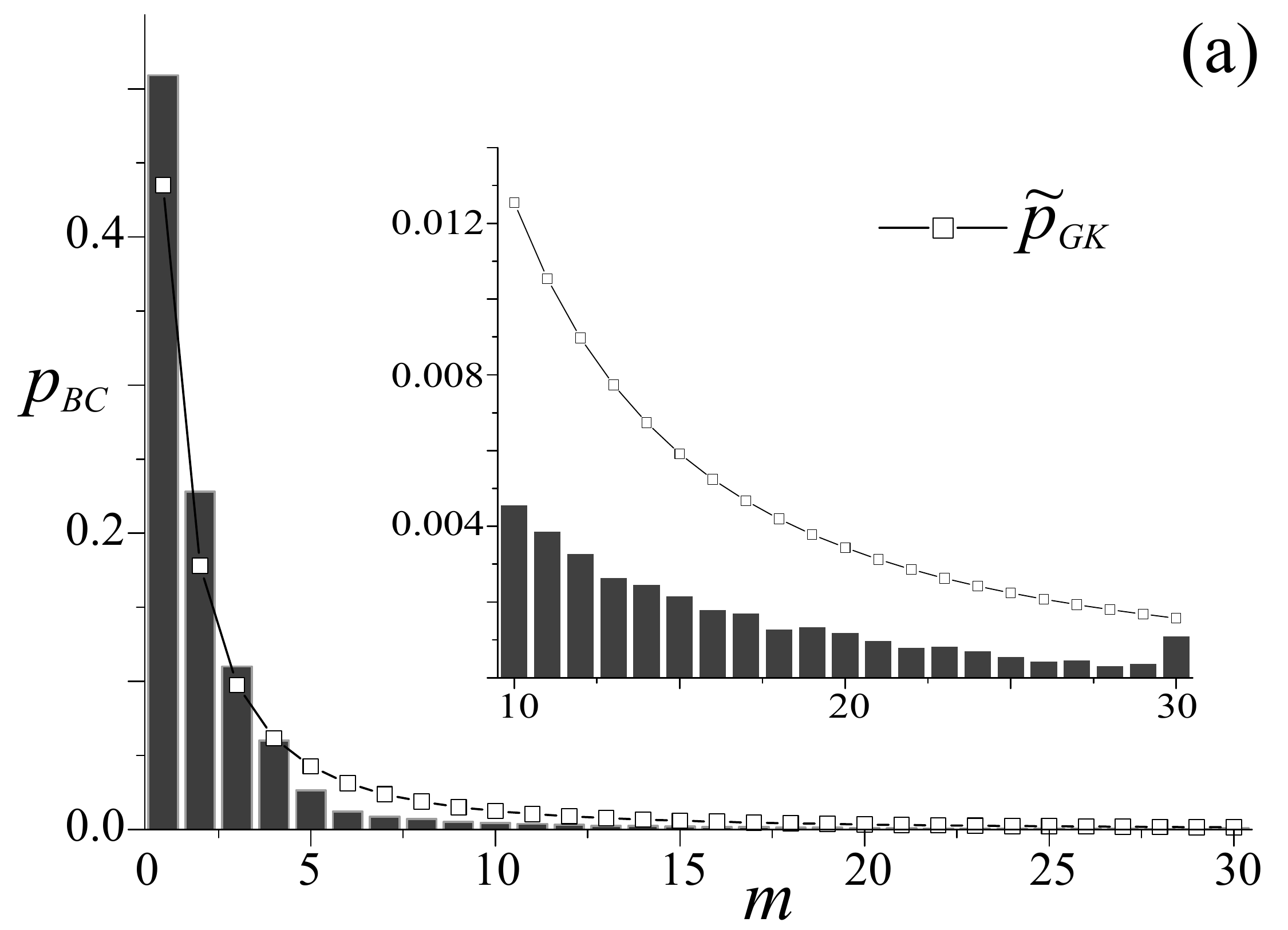}{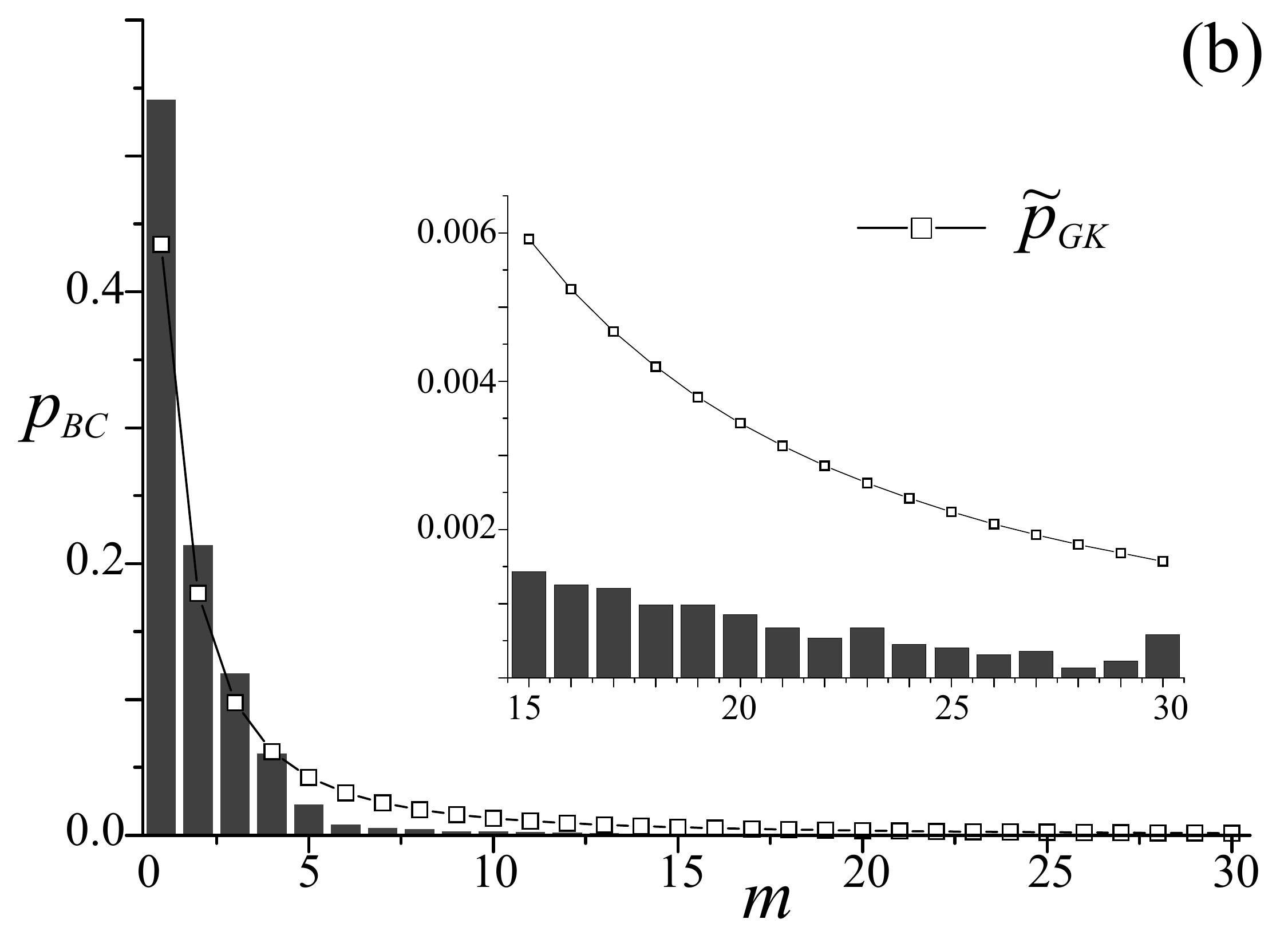}{Distribution of continued fraction elements for the class-one boundary circle $K \in [-0.25,0.75]$ compared with the conditional GK distribution $\tilde{p}_{GK}$ \Eq{GKnormalized}: (a) all coefficients and (b) those for $i>4$.}{CFvsGK}{0.45}
%%%%%

As discussed in \Sec{Building}, we found approximately $2700$
boundary circles of class one for $K \in \left[-0.25,0.75\right]$ and computed, on average, $14$ continued fraction elements for each. The last two coefficients were discarded, as discussed in \Sec{ErrorAnal}. Figure \ref{fig:CFvsGK} shows the empirical probability mass function, $p_{BC}(m)$, the frequency of occurrence of $m\in \bN$ in the resulting list of coefficients $m_i$. Since the  coefficients in any numerical calculation are bounded, $m_i \le m_{max}$, we compare the numerical results to a conditional GK distribution
\begin{equation}\label{eq:GKnormalized}
	\tilde{p}_{GK}(m) =  p_{GK} (m| m < m_{max}) =  Cp_{GK}(m),
\end{equation}
where $C$ is simply the factor that normalizes the distribution up to $m_{max}$
($C=1.05$ for $m_{max} = 30$, as in the figures).

%%%%%
\InsertFigTwo{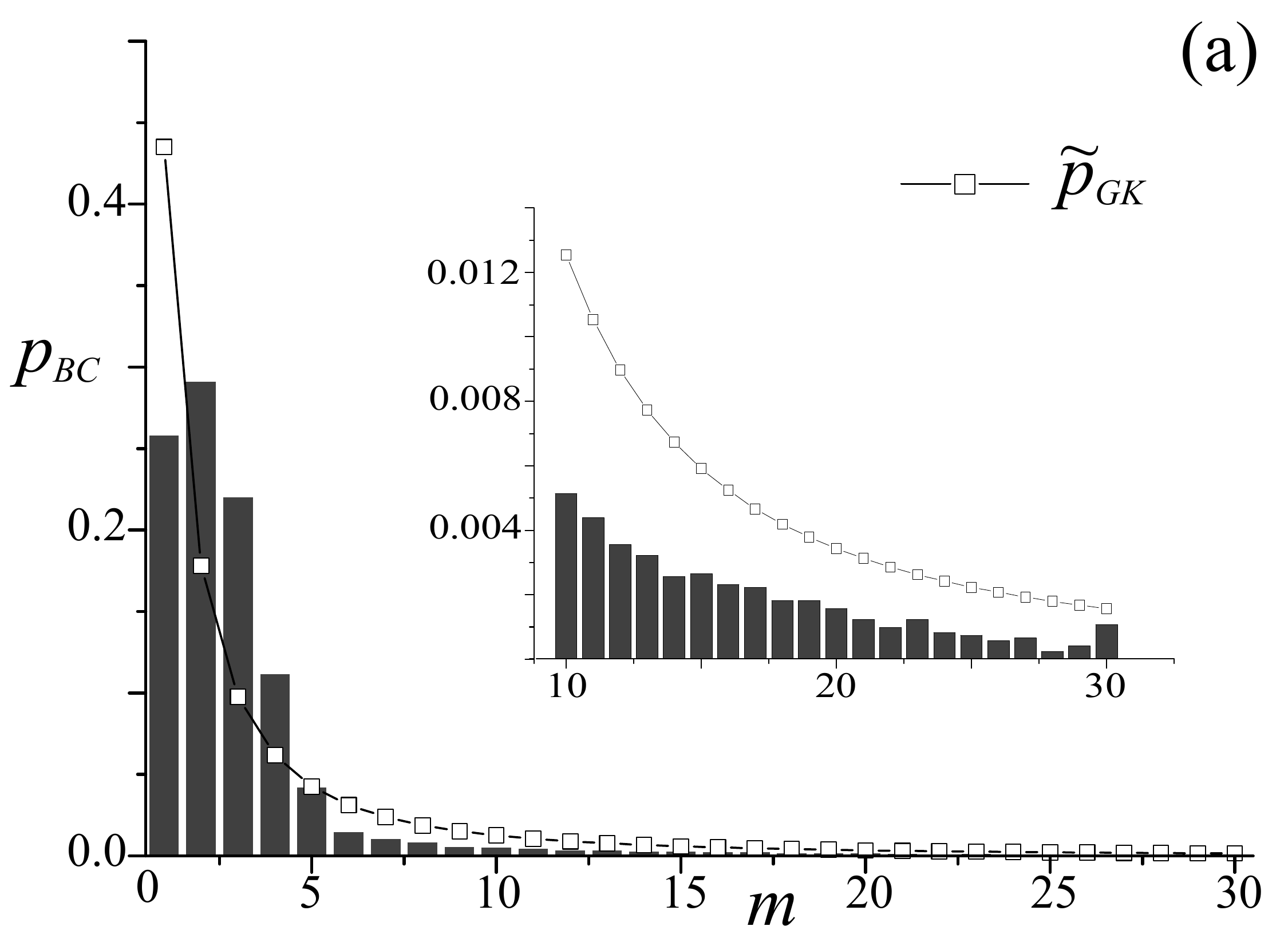}{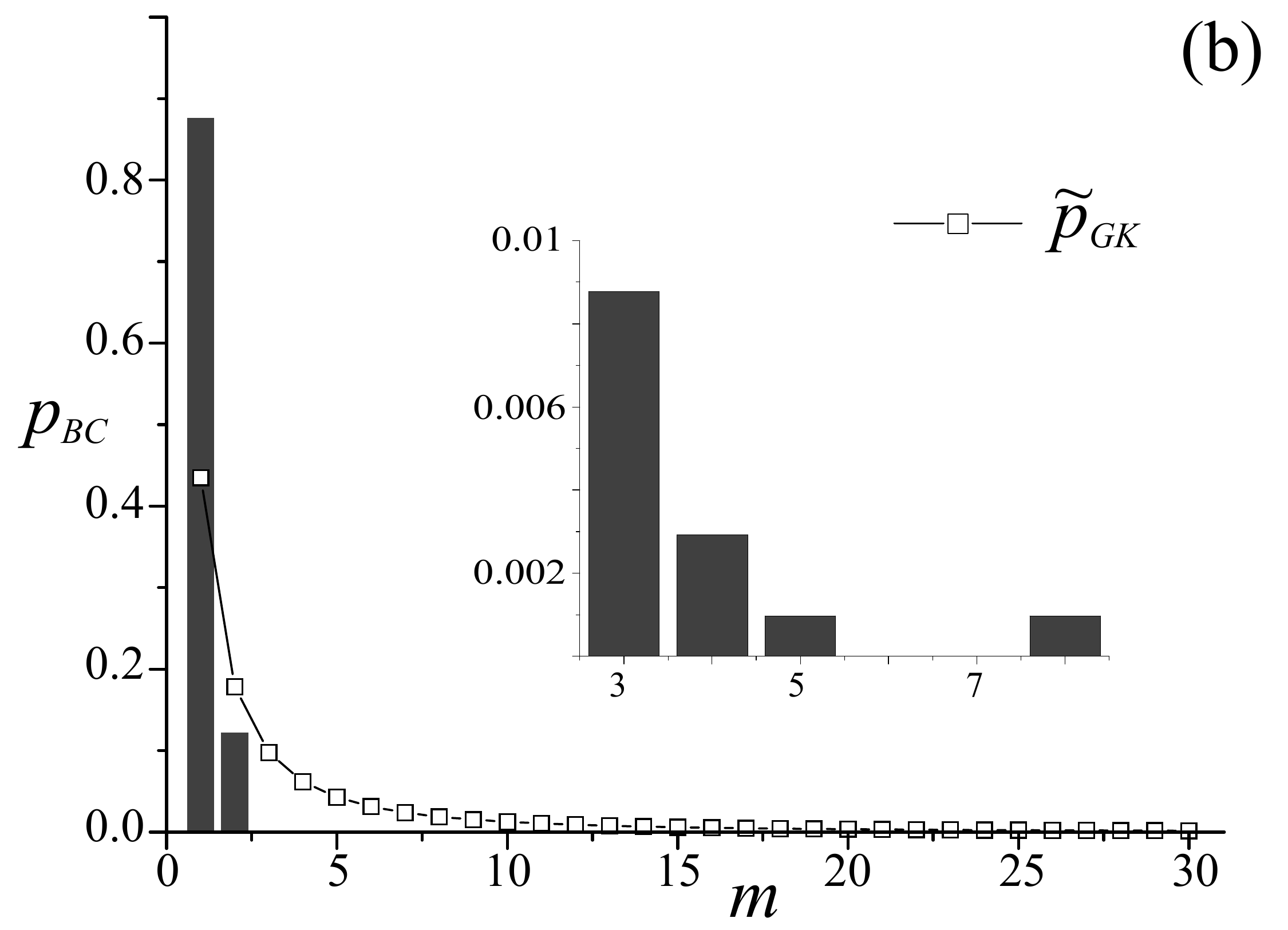}{Distribution of continued fraction elements for the class-one boundary circle for $K \in [-0.25,0.75]$ compared with the conditional distribution $\tilde{p}_{GK}$ \Eq{GKnormalized} using only elements with $i>4$: (a) odd (inner) coefficients and (b) even (outer) coefficients.}{CFvsGKodd_even}{0.45}
%%%%%

Figure \ref{fig:CFvsGK}(a) presents the distribution of
all coefficients for the class-one circle, combining the inner and outer coefficients.
In \Fig{CFvsGK}(b), the same data is shown after discarding the first four coefficients ($m_i$ for $i\le 4$) for each boundary circle, so that details of the global form of the map are de-emphasized. Comparing the two panels of \Fig{CFvsGK} shows that the distribution is not too sensitive to the lower coefficients, though the higher $m$-values occur less frequently in \Fig{CFvsGK}(b).
In both cases, it appears that the distribution is monotonically decreasing at rate that is similar to, though not identical to \Eq{GKnormalized}. Indeed, for $m>5$ the empirical values
of $p_{BC}(m)$ appear to be smaller than $\tilde{p}_{GK}(m)$ by a fixed ratio,
\begin{equation}\label{eq:GKFraction}
	p_{BC}(m) = \alpha  p_{GK}(m).
\end{equation}
We found that for $5<m<30$, $\alpha=0.32\pm0.1$ for the data in \Fig{CFvsGK}(a) and $\alpha=0.21\pm0.04$ for \Fig{CFvsGK}(b).

Figure \ref{fig:CFvsGKodd_even} is the same as \Fig{CFvsGK}(b), but splits the data into inner and outer coefficients. Note that both of these distributions deviate from the GMS results.
For the odd $i$ the distribution still follows \Eq{GKnormalized} for large $m$: for $5<m<30$, $\alpha= 0.39\pm0.08$. The largest element found is $m = 48$ while in GMS the largest value found was $5$.
For the outer coefficients, though we found very few $m_{i}> 3$, some larger coefficients are seen; the largest element found is $m=8$.
Comparing \Fig{CFvsGK}(b) and \Fig{CFvsGKodd_even}(a), we note that if $m_{i}$ with
both $i$ odd and even are taken into account the resulting distribution
for $i > 4$ is closer to $\tilde{p}_{GK}$ than the separate distributions.

These results suggest that the distribution $p_{BC}$ deviates from \Eq{GKnormalized},
and for large $m$ is a fixed fraction as in \Eq{GKFraction}, and thus that
the elements $m_{i}$ for boundary circle rotation numbers are unbounded.

The data in \Fig{CFvsGK} and \Fig{CFvsGKodd_even} are for the class-one boundary circle.
The data for higher class circles is very similar. This data is given in \Tbl{OuterCoef} and \Tbl{InnerCoef}, separated into subsets corresponding to individual states. Thus, for example, the column labeled $S=1$ in these tables corresponds to the distribution of elements for the rotation number of the class-two boundary circle that encircles the elliptic periodic orbit in state $1$. The outer convergents of this circle are represented by the states $S=101^j$. Similarly, the column labeled $10$ corresponds to a class-three boundary circle with outer states $S=1001^j$, etc. The indication is that the distributions are independent of class or site.

%%%%%%%%%%%%%%%%%%%%
\begin{table}[htb]
\begin{tabular}{@{\extracolsep{0.1in}}c|@{\extracolsep{0.1in}}l|@{\extracolsep{0.1in}}l|@{\extracolsep{0.1in}}l|@{\extracolsep{0.1in}}l|@{\extracolsep{0.1in}}l|@{\extracolsep{0.1in}}l|@{\extracolsep{0.1in}}l|@{\extracolsep{0.1in}}l}
  &\multicolumn{8}{c}{$p_{BC}(m)$ (outer coefficients)}\\
\tc{$m$} & \tc{$\emptyset$} & \tc{$1$} & \tc{$10$} & \tc{$11$} & \tc{$100$} & \tc{$101$} & \tc{$110$} & \multicolumn{1}{c}{$111$}\\
\hline
$1$ & $0.87624$ & $0.86467$ & $0.92444$ & $0.90103$ & $0.92857$ & $0.97273$ & $0.94684$ & $0.92239$\\
$2$ & $0.12239$ & $0.12911$ & $0.07556$ & $0.09616$ & $0.07143$ & $0.02727$ & $0.05316$ & $0.07761$\\
$3$ & $0.00088$ & $0.00329$ &   & $0.00094$ &   &   &   &  \\
$4$ & $0.00029$ & $0.00183$ &   & $0.00187$ &   &   &   &  \\
$5$ & $0.00010$ & $0.00037$ &   &   &   &   &   &  \\
$6$ &   & $0.00037$ &   &   &   &   &   &  \\
$7$ &   & $0.00037$ &   &   &   &   &   &  \\
$8$ & $0.00010$ &   &   &   &   &   &   &  \\
\hline
$Total\, \#$ & $10246$ & $2734$ & $675$ & $3203$ & $14$ & $110$ & $301$ & $670$\\
\end{tabular}
\caption{Probabilities of appearance of an entry $m$ as an outer (even) coefficient of the continued fraction expansion of a boundary circle of the specific state. The first four ($i\le4$) and the last two coefficients for each continued fraction have been discarded.
\label{tbl:OuterCoef}}
\end{table}
%%%%%%%%%%%%%

%%%%%%%%%%%
\begin{table}[htb]
\begin{tabular}{@{\extracolsep{0.1in}}c|@{\extracolsep{0.1in}}l|@{\extracolsep{0.1in}}l|@{\extracolsep{0.1in}}l|@{\extracolsep{0.1in}}l|@{\extracolsep{0.1in}}l|@{\extracolsep{0.1in}}l|@{\extracolsep{0.1in}}l|@{\extracolsep{0.1in}}l}

   &\multicolumn{8}{c}{$p_{BC}(m)$ (inner coefficients)}\\
\tc{$m$} & \tc{$\emptyset$} & \tc{$1$} & \tc{$10$} & \tc{$11$} & \tc{$100$} & \tc{$101$} & \tc{$110$} & \multicolumn{1}{c}{$111$}\\
\hline
$1$ & $0.25776$ & $0.25926$ & $0.2779$ & $0.31711$ & $0.38554$ & $0.39614$ & $0.3467$ & $0.34167$\\
$2$ & $0.29086$ & $0.30047$ & $0.28322$ & $0.31548$ & $0.25301$ & $0.30918$ & $0.26179$ & $0.28653$\\
$3$ & $0.21986$ & $0.23670$ & $0.21792$ & $0.21550$ & $0.13253$ & $0.15459$ & $0.19104$ & $0.21319$\\
$4$ & $0.11146$ & $0.12448$ & $0.11314$ & $0.10100$ & $0.07229$ & $0.07730$ & $0.12028$ & $0.10176$\\
$5$ & $0.04195$ & $0.04623$ & $0.06530$ & $0.04110$ & $0.07229$ & $0.03865$ & $0.04599$ & $0.05060$\\
$6$ & $0.01465$ & $0.01058$ & $0.02202$ & $0.00593$ & $0.01205$ & $0.01449$ & $0.01533$ & $0.00512$\\
$7$ & $0.01026$ & $0.00501$ & $0.00987$ & $0.00225$ & $0.01205$ & $0.00966$ & $0.00708$ & $0.00114$\\
$8$ & $0.00836$ & $0.00501$ & $0.00228$ & $0.00041$ & $0.01205$ &   & $0.00472$ &  \\
$9$ & $0.00554$ & $0.00223$ & $0.00152$ & $0.00061$ & $0.02410$ &   & $0.00354$ &  \\
$10$ & $0.00513$ & $0.00278$ & $0.00152$ & $0.00020$ & $0.01205$ &   &   &  \\
$11$ & $0.00439$ & $0.00167$ & $0.00152$ &   &   &   &   &  \\
$12$ & $0.00356$ & $0.00084$ & $0.00152$ &   &   &   & $0.00236$ &  \\
$13$ & $0.00323$ & $0.00139$ &   & $0.00020$ &   &   &   &  \\
$14$ & $0.00257$ & $0.00111$ & $0.00076$ & $0.00020$ &   &   & $0.00118$ &  \\
$15$ & $0.00265$ & $0.00056$ &   &   &   &   &   &  \\
$\ge 16$ &$0.01778$ & $0.00167$  &$0.00152$ &   &   $0.01205$ &  & &\\
%$16$ & $0.00232$ & $0.00111$ &   &   &   &   &   &  \\
%$17$ & $0.00223$ &   		 & $0.00076$ &   &   &   &   &  \\
%$18$ & $0.00182$ &   		 &   &   &   &   &   &  \\
%$19$ & $0.00182$ &   		 & $0.00076$ &   &   &   &   &  \\
%$20$ & $0.00157$ &   		 &   		 &   & $0.01205$ &   &   &  \\
%$21$ & $0.00124$ &   		 &   &   &   &   &   &  \\
%$22$ & $0.00099$ & $0.00028$ &   &   &   &   &   &  \\
%$23$ & $0.00124$ &  		 &   &   &   &   &   &  \\
%$24$ & $0.00083$ & $0.00028$ &   &   &   &   &   &  \\
%$25$ & $0.00074$ &   &   &   &   &   &   &  \\
%$26$ & $0.00058$ &   &   &   &   &   &   &  \\
%$27$ & $0.00066$ &   &   &   &   &   &   &  \\
%$28$ & $0.00025$ &   &   &   &   &   &   &  \\
%$29$ & $0.00041$ &   &   &   &   &   &   &  \\
%$30$ & $0.00108$ &   &   &   &   &   &   &  \\
\hline
$Total\, \#$ & $12085$ & $3591$ & $1317$ & $4891$ & $83$ & $414$ & $848$ & $1759$\\

\end{tabular}
\caption{Probabilities of appearance of an entry $m$
as an inner (odd) coefficient of the continued fraction expansion
of a boundary circle of the specific state. The first four ($i\le4$)
and the last two coefficients for each continued fraction have been discarded.
\label{tbl:InnerCoef}}
\end{table}

We turn now to a heuristic explanation of our calculations.
KAM theory implies that smooth, Diophantine invariant circles are structurally stable. Even
though for a subclass of Diophantine numbers---those for which there is a sufficiently small $C>0$ such that there is no solution to $|\omega-\tfrac{p}{q}|<\tfrac{C}{q^2}$ for integers $(p,q)$ with $q>0$---there is a bound on the continued fraction elements \cite[Thm. 23]{Khinchin1964}. This bound is not uniform in $C$, thus for a random selection of Diophantine numbers, there is no uniform bound on the elements. The outer coefficients correspond to periodic orbits on the chaotic side of the boundary circle: from this side the circle appears isolated. Critical circles with noble rotation numbers (numbers for which the elements in the tail of the continued fraction are all $1$'s) were found in \cite{MacKay1983,Mackay1988,Mackay1989} using the renormalization methods introduced by MacKay \cite{MacKay1983} and further developed in \cite{Mackay1988,Mackay1989}. These are isolated from both sides. Following the more general argument of GMS, we expect that the outer (even $i$) coefficients will eventually have this property. Even the smaller $i$ coefficients show the strong influence of the robustness of noble rotation numbers: $m_i = 1$ occurs about $85\%$ of the time. The inner coefficients should have no such restriction: the $m_i$ should satisfy the GK distribution. However, the small $m_i$ coefficients are still influenced by the \emph{nobility} of the robust circles, and so their occurrence exceeds that  predicted by \Eq{Gauss-Kuzmin}. Combining these ideas implies that distribution for the larger values of $m$ should be a fixed fraction of the GK distribution, as \Eq{GKFraction}.
Finally for small $i$, the $m_i$ are influenced by the actual value of the winding number as a function of $K$, \Eq{ResidueRotation}, and thus will not be uniformly distributed.
Nevertheless, we found that the  distributions for the larger, $i>4$, coefficients are similar to those found when the contributions of all $i$ are taken into account.

%%%%%%%%%%%%%%%%%
%% Distribution of Flux
%%%%%%%%%%%%%%%%%
\section{ Distribution of Flux Ratios}\label{sec:Flux}

In this section we compute the distribution of the flux ratios $w_S^{(i)}$, defined by \Eq{PeriodicFluxRatio},
associated with the states on the Markov tree.
Since these ratios are all positive, but vary over several orders of magnitude, we
will analyze their log-distribution, or equivalently the distribution of
\begin{equation}\label{eq:vdef}
       v^{(i)}_{S}=-\ln w^{(i)}_{S} .
\end{equation}
We chose the minus sign above so that $v_S^{(i)} >0$ when the flux ratio is smaller than one, which, as noted in \Sec{ErrorAnal}, is expected since areas should decrease when level or class is incremented.
In particular we will study the distribution for the right-going steps
(recall \Fig{SelfSimilarTree}), corresponding
to incrementing the level, $v_S^{(1)} = v_S^{level}$ and the left-going steps, $v_S^{(0)} = v_S^{class}$ corresponding to incrementing the class, separately.
We will use states with up to four elements, so given that \Eq{PeriodicFluxRatio} is a ratio of daughter to parent flux, there will be seven parent states for each level and class transition,
\begin{equation}\label{eq:SevenStates}
	 \{1, 10, 11, 100, 101, 111\} .
\end{equation}
Thus, $v_S^{level}$ is computed using \Eq{vdef} with a state $S$ selected from \Eq{SevenStates} and a daughter $S1$, and $v_S^{class}$ is computed for the same states, but with a daughter $S0$.

The computed distributions $p(v^{level})$ and $p(v^{class})$ are presented in \Fig{Vdistributions} for data from two subsets of the full range $K \in [-0.25,0.75]$ combining ratios from all of the parent states in \Eq{SevenStates}.
In both cases, the distributions are similar in the different intervals of $K$.

%%%%%
\InsertFigTwo{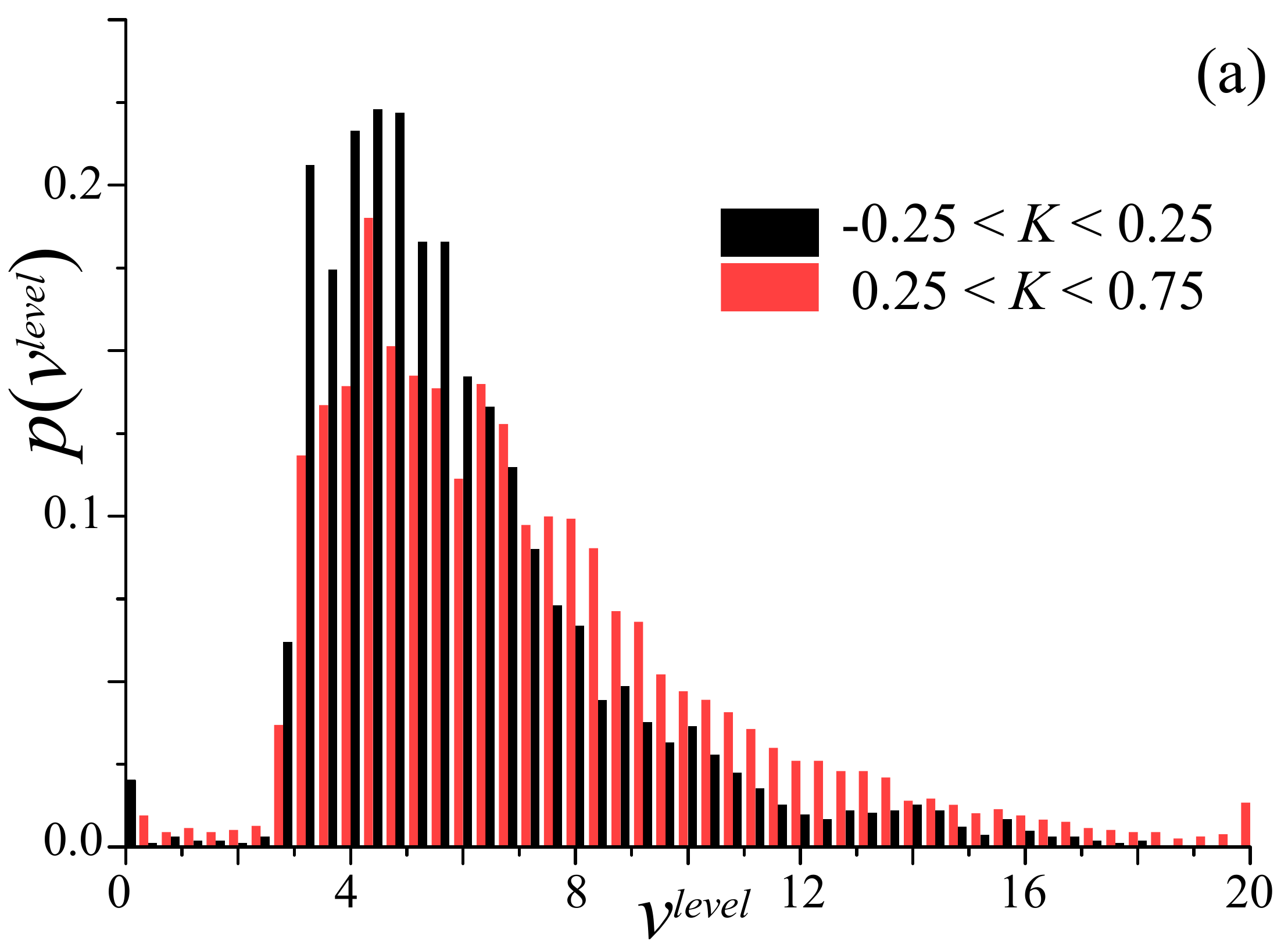}{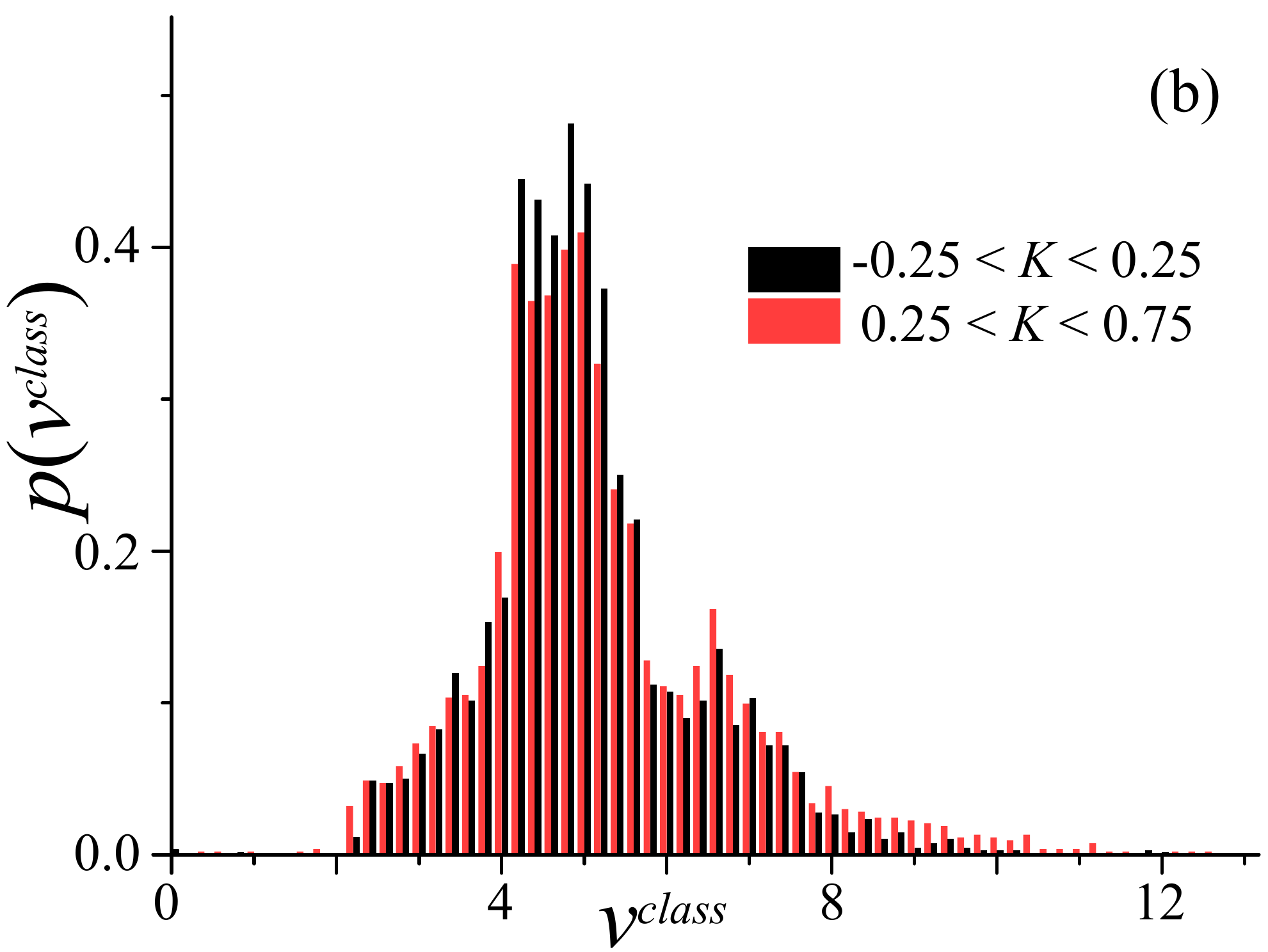}{(Color online) Probability
distributions of the log-flux ratio \Eq{vdef}. (a) Distribution of $v^{level}$, corresponding to right-going transitions $S\to S1$ for $S$ in \Eq{SevenStates} using bins of size $0.4$. (b) Distribution of $v^{class}$, corresponding to left-going transitions, $S \to S0$ using bins of size $0.2$. The shades (black and red) show two ranges of $K$ as indicated.}{Vdistributions}{0.45}
%%%%%

We now fit two model distribution families using the binned-likelihood method to determine the best values of the parameters to maximize the log-likelihood for the data.
The binned-likelihood method was implemented by the framework ``Root'' \cite{BrunRene}.
The first model is a three-Gaussian mixture model
\begin{equation}\label{eq:Gaussian}
       p_{G}(x)=\sum_{j=1}^3
\alpha_{j}e^{-\frac{(x-\mu_{j})^{2}}{2\beta_{j}^{2}}}
\end{equation}
with parameters $\alpha_{j},\beta_{j},\mu_{j}$, of which eight are independent upon normalization.\footnote
%%%%
{We also attempted fits with one or two Gaussian mixtures, but these fit the data poorly.}
%%%%
The second is a shifted Gamma distribution,
\begin{equation}\label{eq:Gamma}
       p_{\Gamma}(x)=
               \left\{ \begin{array}{ll}
                       \frac{(x-\mu_1)^{\alpha_1-1} e^{-\frac{x-\mu_1}{\beta_{1}}}}
                         {\beta_{1}^{\alpha_1}\Gamma(\alpha_1)} & x > \mu_1\\
               0 &  x < \mu_1 \\
               \end{array} \right. ,
\end{equation}
with the three parameters $\alpha_1,\beta_1,\mu_1$.
For the fits, as in \Fig{Vdistributions}, a bin size of $\Delta v = 0.2$ was used for $v^{class}$ and of $0.4$ for $v^{level}$.
Comparisons of the raw event count distributions with the best fits are presented in \Fig{vFits}. Since the distributions in the figures are event counts, they are not normalized, so the Gamma distribution is multiplied by a constant $C$, and the Gaussian mixture has unnormalized amplitudes $\alpha_i$. The values of the best fit parameters are presented in \Tbl{vFitParameters}.

%%%%%
\InsertFigTwo{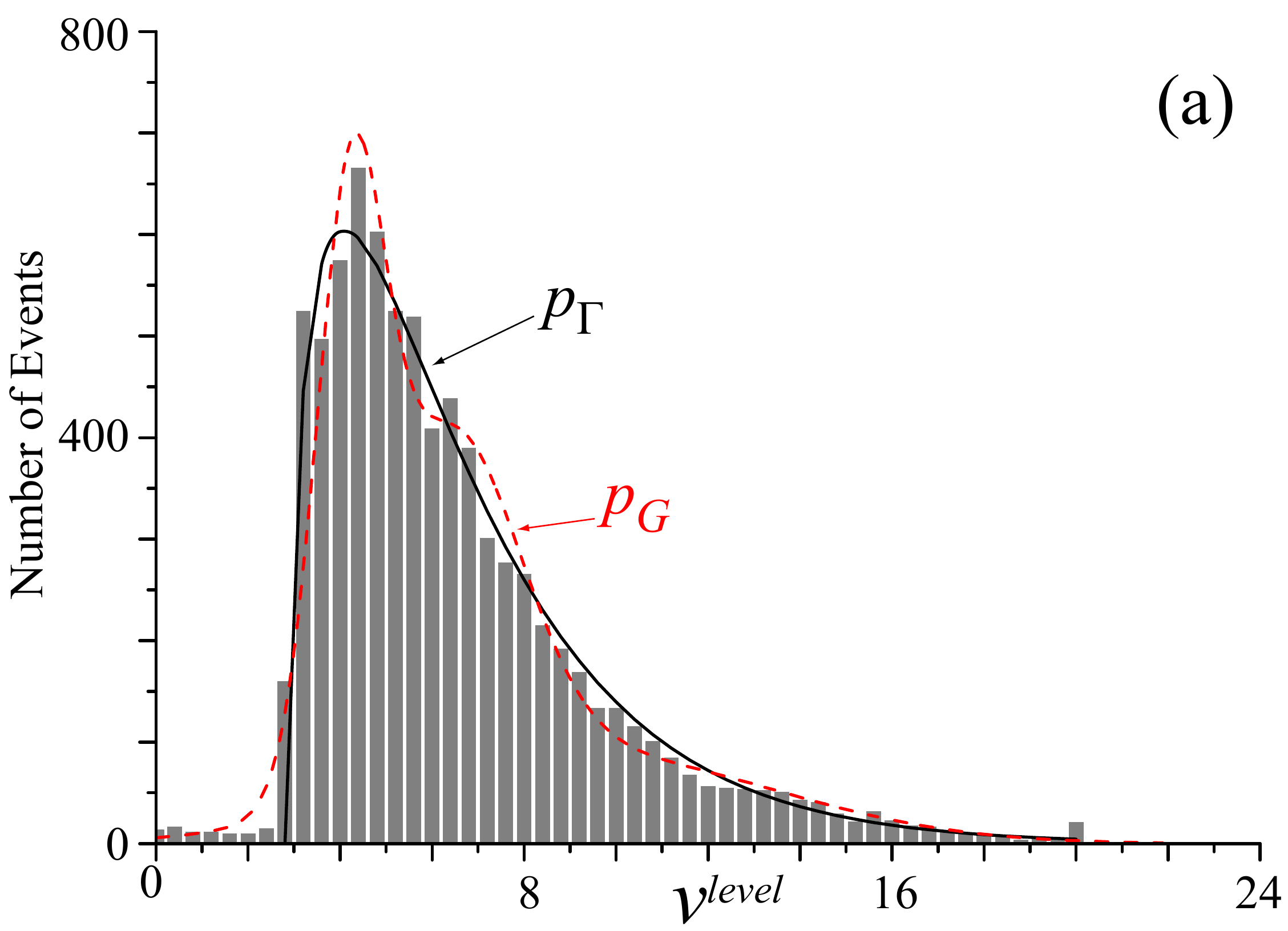}{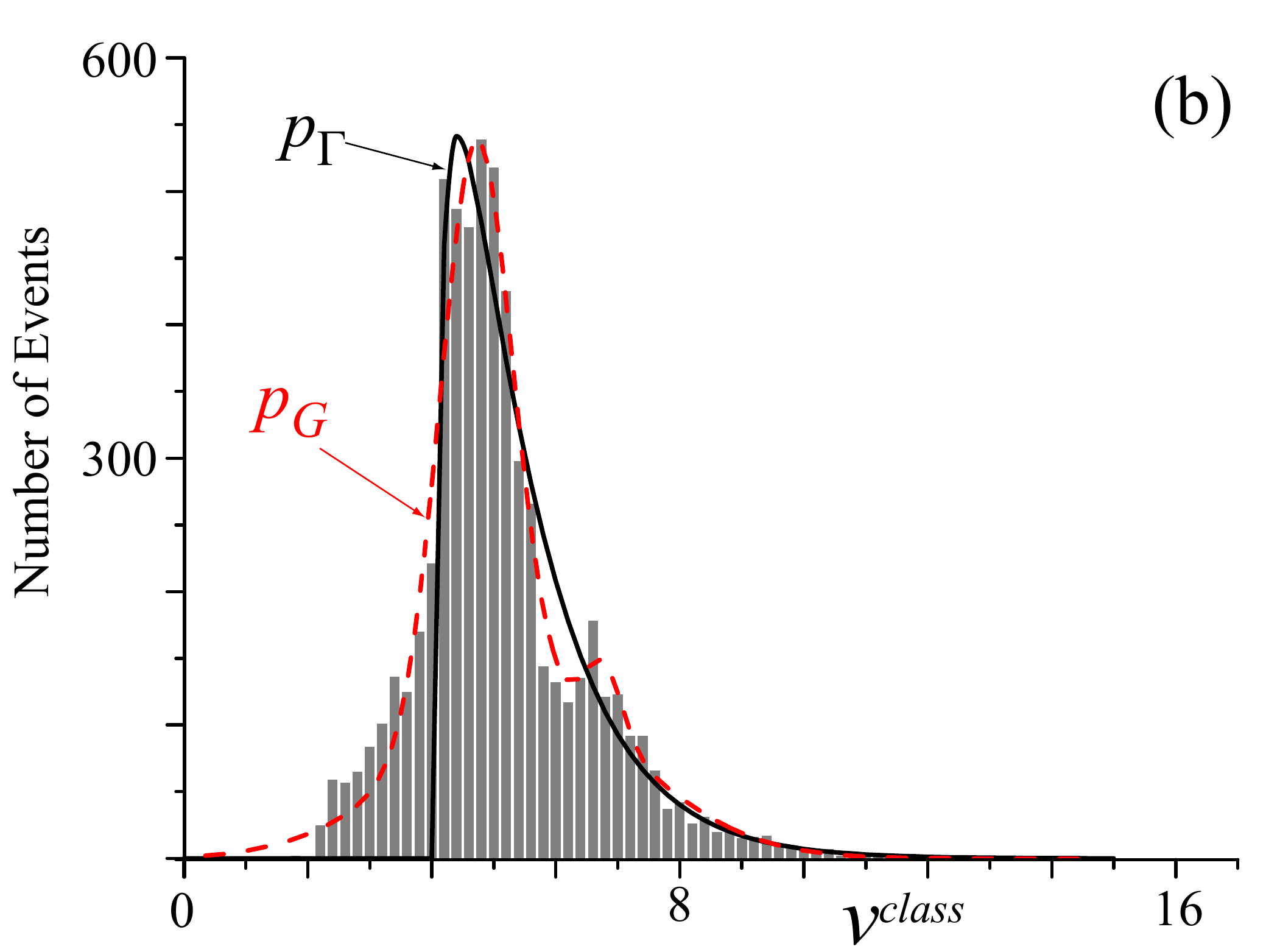}{(Color online) Event distributions for $v^{level}$ (a) and $v^{class}$ (b), and fits to the Gaussian mixture model \Eq{Gaussian} (red-dashed line) and the Gamma distribution \Eq{Gamma} (black-solid line).
Parameter values used in the fits are given in \Tbl{vFitParameters}.}{vFits}{0.45}
%%%%%

%%%%%%%%%%%%%%
\begin{table}[H]
\centering
\begin{tabular}{c|c|c|c|c}
                        & \multicolumn{2}{c|}{$v^{level}$} &
\multicolumn{2}{c}{$v^{class}$}\\
Parameter & $p_\Gamma$ & $p_G$ & $p_\Gamma$ & $p_G$\\
\hline
{$\alpha_{1}$} 	& {$1.5\pm0.2$} & {$527\pm22$} &   {$1.26\pm0.06$} 	& {$444\pm13$}\\
{$\alpha_{2}$} 	&  				& {$342\pm15$} &  					& {$114\pm13$}\\
{$\alpha_{3}$} 	&  				& {$86\pm5.5$} &  					& {$56\pm10$}\\

{$\mu_{1}$} & {$2.92\pm0.03$} 	& {$4.21\pm0.04$} 	& {$4.12\pm0.06$}& {$4.71\pm0.02$}\\
{$\mu_{2}$} &  					& {$6.37\pm0.09$}	&  				 & {$5.52\pm0.05$}\\
{$\mu_{3}$} &  					& {$9.4\pm0.2$} 	&  				 & {$6.76\pm0.06$}\\

{$\beta_{1}$} & {$2.6\pm0.2$} 	& {$0.73\pm0.03$} 	& {$1.09\pm0.04$}& {$0.57\pm0.03$}\\
{$\beta_{2}$} &  				& {$1.52\pm0.05$} 	&  				 & {$1.83\pm0.05$}\\
{$\beta_{3}$} &  				& {$4.1\pm0.1$}	&  				 	 & {$0.3\pm0.07$}\\

{$C$} 		  & {$3122$} 		&  					& {$988$} 		 & \\
\hline
{$f$}		  & {$0.12$} 		& {$0.015$} 		& {$0.22$} 		 & {$0.05$}\\
\end{tabular}
\caption{The fitting parameters to the Gaussian mixture
\Eq{Gaussian} and Gamma \Eq{Gamma} distributions for the computed
values of $v^{level}$ and $v^{class}$. The last
row gives the fraction $f$ of $p_G$ or $p_\Gamma$ in the tails, \Eq{fraction}.}
\label{tbl:vFitParameters}
\end{table}
%%%%%%%%%%%%%%%%

%%%%%
\InsertFigTwo{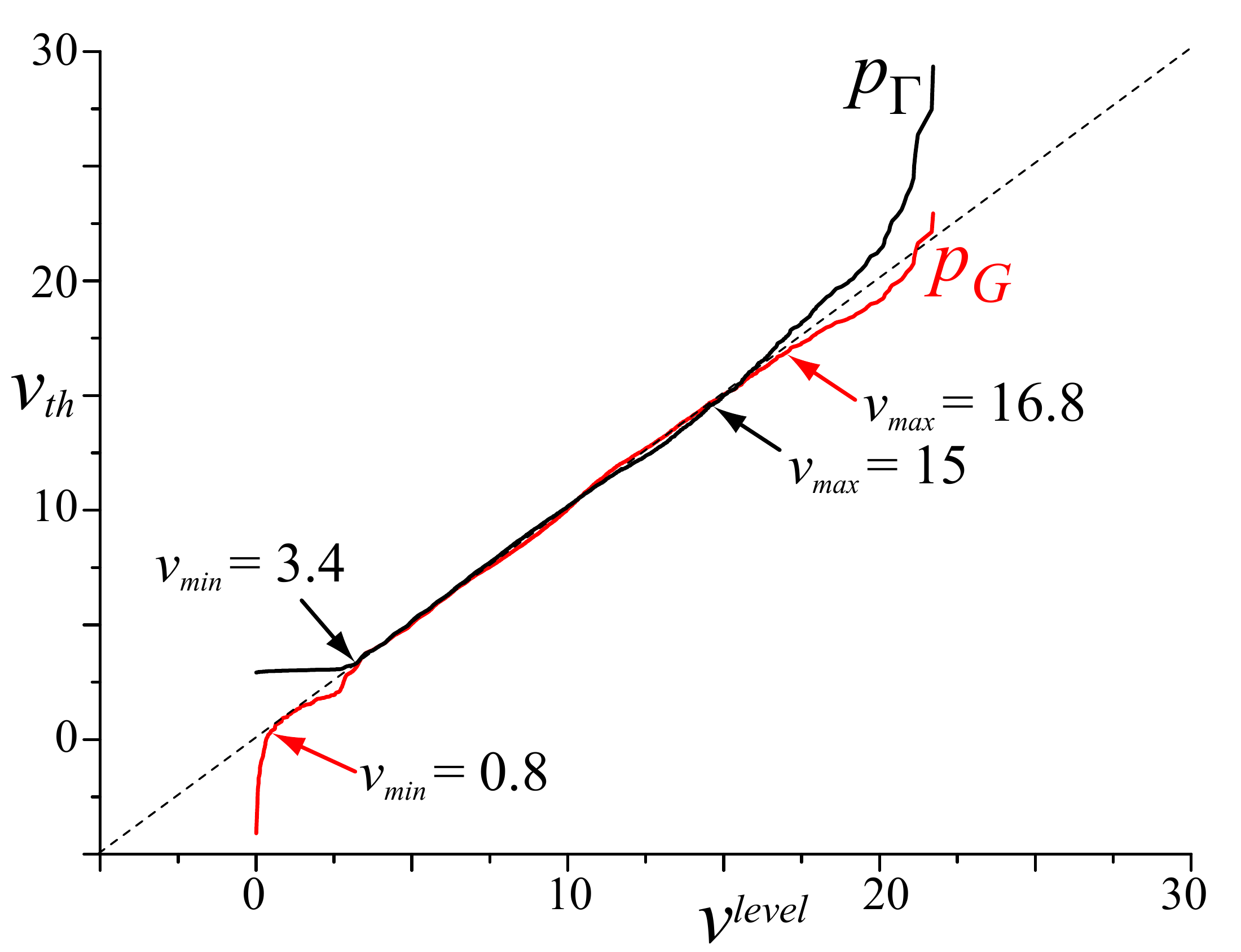}{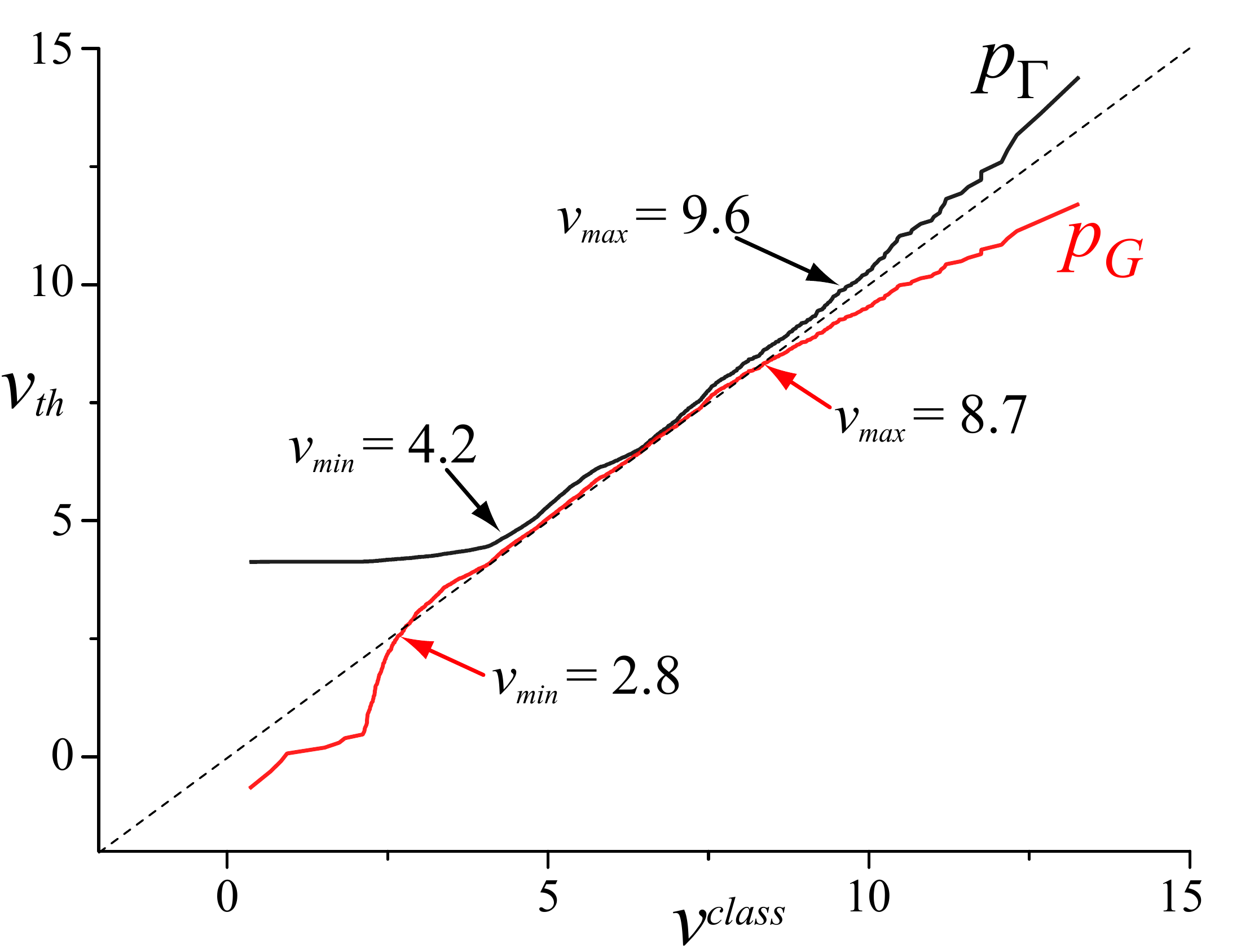}{(Color online) Quantile-Quantile plots of the
numerical data for
$v^{level}$ (a) and $v^{class}$ (b) compared with the fits to the
Gaussian mixture model with three components, and the Gamma
distribution.}{QQplots}{0.45}
%%%%%

In order to evaluate the quality of the fit we use the quantile-quantile plots
presented in \Fig{QQplots}. For each quantile $v$ in the data we find the quantile $v_{th}$ so that the
cumulative distributions match: $P_{th}(v_{th})=P_{dat}(v)$,
and then plot $v_{th}$ as a function of $v$.
Here $P_{th}$ is taken to be $P_{G}$ or $P_\Gamma$, the fitted analytical cumulative distribution and $P_{dat}$ is the numerical one.
From \Fig{QQplots}  we see that for all cases, there is a range $v_{min} < v < v_{max}$ over which the fit is good, but that in the tails, the fits deviate from the data.
The weight of the regions outside $[v_{min},v_{max}]$ is the quantity
\begin{equation}\label{eq:fraction}
	f=P_{th}(v_{min})+1-P_{th}(v_{max}).
\end{equation}
For the Gaussian mixture model, $f \le 0.05$ for both cases, indicating the high quality
of this fit. For the Gamma distribution, $f$ is larger, though it seems
acceptable as a fit to the distribution of $v^{level}$, this is confirmed visually in \Fig{vFits}. One has to keep in mind that Gaussian mixture model has eight parameters, many more than the three of the Gamma distribution.

Finally, we will compare the fitted distribution to a smoothed version of the data
using the characteristic function to do the smoothing.
As usual, to extract the distribution of the $v$, we have collected the data into bins and fit the distribution to the occupation numbers of the bins. This partition into bins is somewhat
arbitrary and may lead to details that are incorrect
on fine scales. To gain control of the various scales we will use
the characteristic function
\begin{equation}\label{eq:Characteristic}
       f_C(k)=\left\langle e^{-ikz}\right\rangle =\intop dz \, p(z)e^{-ikz} .
\end{equation}
The Fourier transformation will be computed using bins but it will interpolate different scales. Here we work with standardized set of $J$ bins uniform on $[v^{low},v^{high}]$.
If $X_{j}$ is the number of $v$'s in the $j^{th}$ bin, then its weight is $p_{j}=\frac{X_{j}}{C}$, with normalization constant $C$.
The integral for the characteristic function \Eq{Characteristic} is computed using the trapezoidal rule upon transforming the integration variable to have average $0$ and standard deviation $1$. Transforming back we obtain a smoothed distribution $p_C(v)$ of the original data. The width of the bins is chosen to be $\Delta v=0.2$ for both $v^{level}$ and
for $v^{class}$.
The smoothed distribution $p_C(v)$ gives
indeed a faithful representation of the original data according to a
Kolmogorov-Smirnov test: the null hypothesis that the distributions are the same has significance values $1$ for levels and $0.88$ for classes. A comparison of the smoothed distribution with the fits $p_G$ and $p_\Gamma$ are shown in \Fig{vFitswithCF}.

%%%%%%
\InsertFigTwo{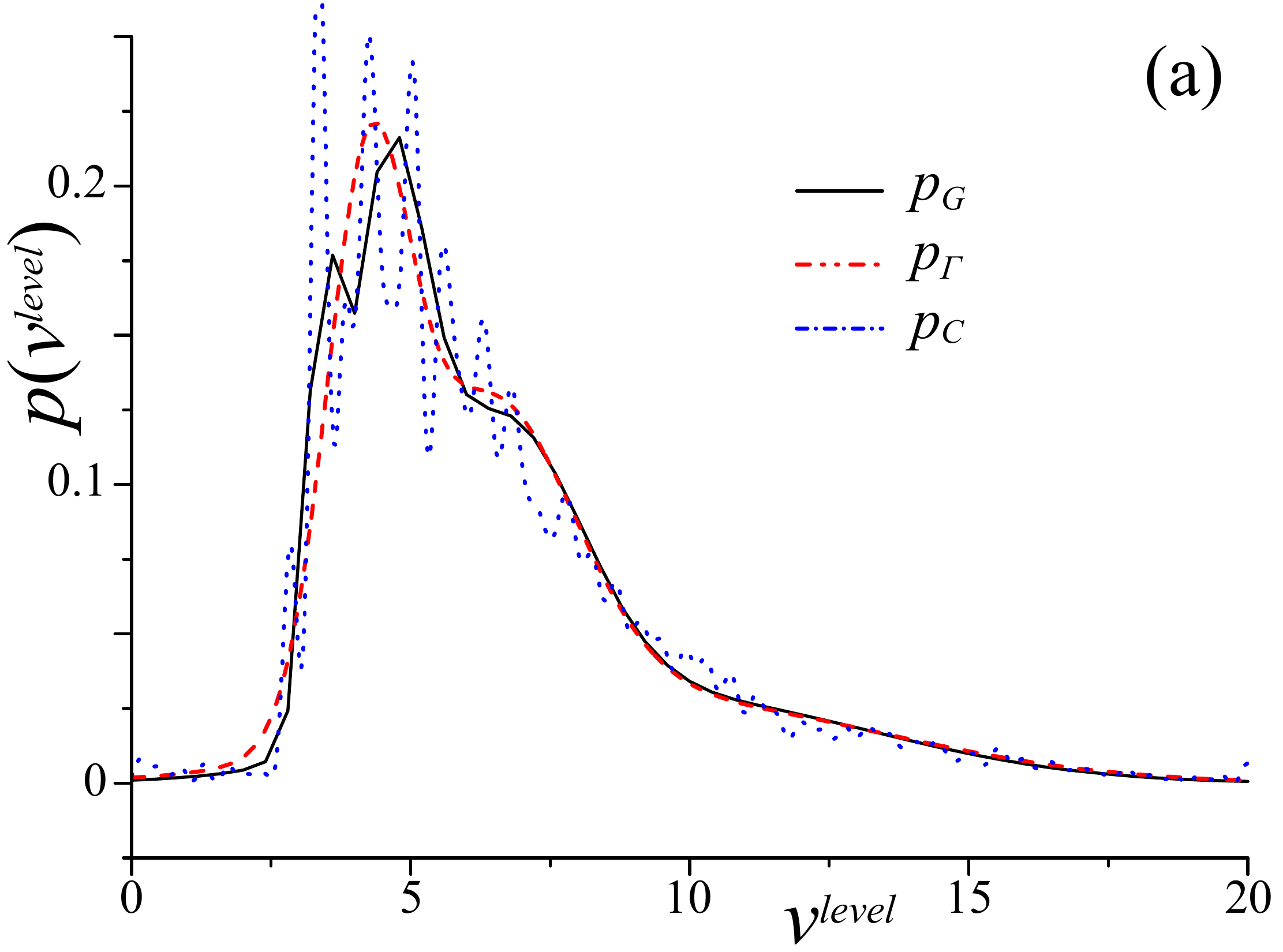}{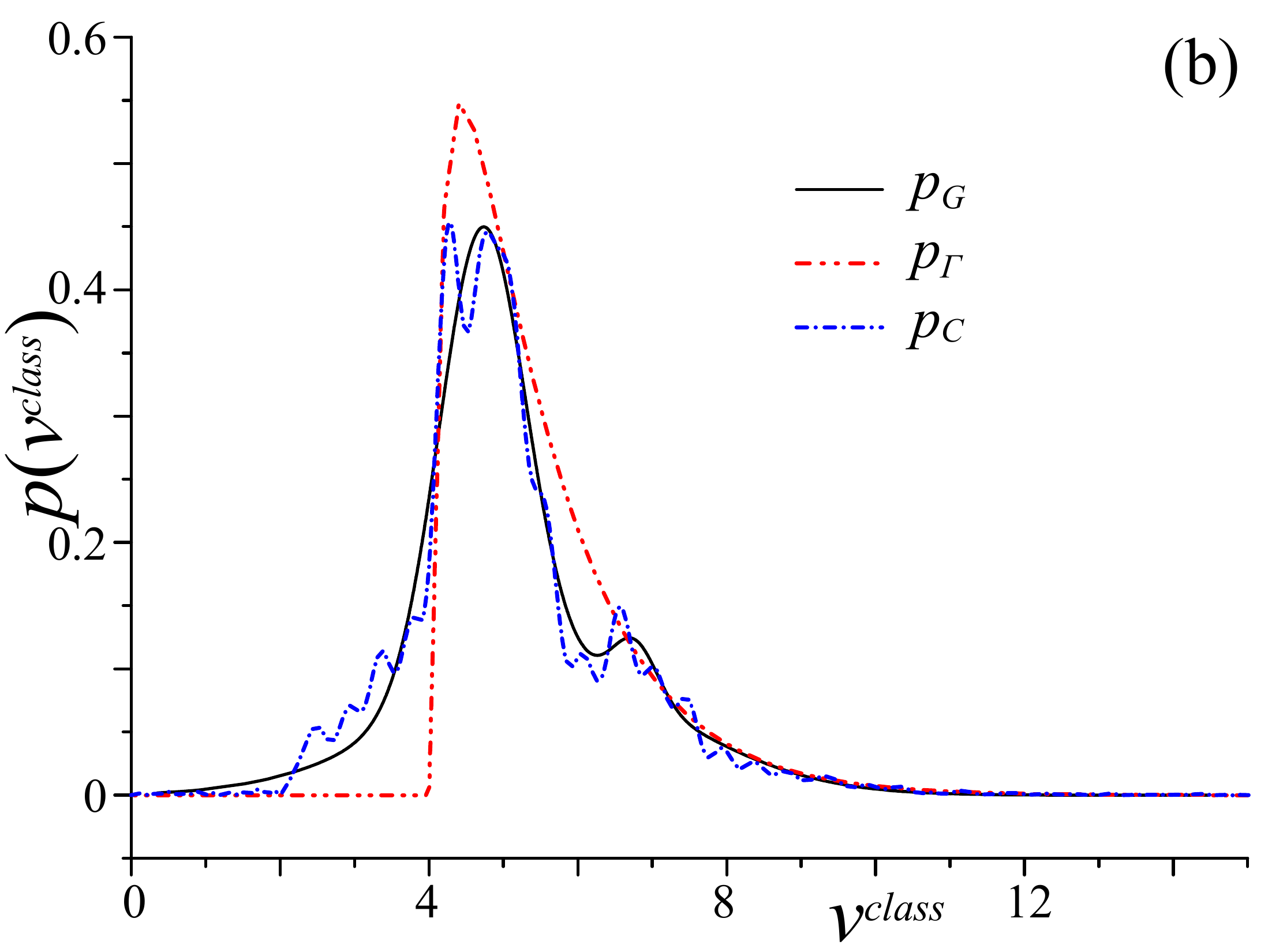}{(Color online) Fitted Gaussian mixture and shifted Gamma distributions using the parameters in \Tbl{vFitParameters}, compared to the smoothed characteristic function fit $p_C$ for (a) levels and (b) classes.}{vFitswithCF}{0.45}
%%%%%%

Finally, we computed distributions of the flux ratios $v_S^{(i)}$ for the individual states \Eq{SevenStates} and found that their shape appears to be independent of state. This indicates that the distributions shown in \Fig{vFitswithCF} appear to be universal for different levels of the island-around-islands hierarchy.

%%%%%%%%%%%%%%%%%
%% Summary
%%%%%%%%%%%%%%%%%
\section{Summary}\label{sec:Summary}

We have presented some statistical properties of boundary circle rotation numbers and flux ratios for the H\'enon map \Eq{T} over the full range of $K$ for which it has an elliptic fixed point. These quantities play an important role in the Markov tree model for transport as formulated by Meiss and Ott \cite{Meiss1986a}.

We analyzed the distribution of the elements $m_i$ of the continued
fraction for boundary circle rotation numbers and found, in contrast
to the results of Greene, MacKay and Stark \cite{Greene1986},
strong evidence that the inner coefficients (odd $i$) are unbounded.  The
distribution was contrasted to the Gauss-Kuzmin \Eq{Gauss-Kuzmin} distribution, which applies
to a uniform distribution of irrational numbers. We found that the fraction of elements taking small values is larger than expected
from Gauss-Kuzmin and conversely that large elements occur with smaller probability. Moreover we have some indications that for $m \ge 5$, the distribution is a fixed fraction of that expected from Gauss-Kuzmin. The evidence presented indicates that the empirical distribution is universal over different island hierarchies in the phase space (see \Tbl{InnerCoef}).
An analytical understanding of these results awaits future studies.

We studied the distribution of flux ratios that determine the decrease in flow through phase space upon moving to smaller structures corresponding to islands-around islands (classes) or being trapped nearer to a boundary circle (levels).
The numerical results indicate that there are two universal distributions, one for classes and one for levels, independent of the parent state and of the value of the parameter $K$. From \Tbl{vFitParameters} and \Fig{vFits}, it is clear that the distributions for levels and classes are significantly different.
Visually it appears that the distribution of flux ratios for levels is a log-Gamma distribution, while that for classes is a superposition of log-Gaussian distributions.

The existence of such distributions is a basic assumption of various
recent studies on the universality of algebraic decay of correlations and Poincar\'e recurrence distributions for area-preserving maps \cite{Cristadoro2008,Ceder2013}.
In the future, we plan to compute the distribution of the area ratios, \Eq{AreaRatio}, and hope to verify that the distributions that we have observed here also apply to other maps, such as Chirikov's standard map. We also leave the study of the validity of the Markov approximation to future studies.

\begin{acknowledgments}
The work was supported in part by the Israel Science Foundation (ISF) grant number
1028/12, by the US-Israel Binational Science Foundation (BSF) grant
number 2010132, by the Shlomo Kaplansky academic chair,
and by the US National Science Foundation (NSF) grant DMS-1211350.
We would like to thank Roland Ketzmerick, William Kleiber, Manual Lladser, Yoran Rozen,
and Ed Ott for fruitful discussions.
\end{acknowledgments}

\bibliographystyle{unsrtnat}
\bibliography{StatChar}

\end{document}